\documentclass[aps,prd,twocolumn,superscriptaddress,nofootinbib,longbibliography]{revtex4-2}

\usepackage{graphicx}
\usepackage{amsmath,amssymb,bm}
\usepackage{booktabs}
\usepackage{xcolor}
\usepackage{flafter}
\usepackage{float}
\usepackage{placeins}
\usepackage{hyperref}

\makeatletter

\newenvironment{resultfigurewide}{\begin{figure*}[tp]\centering}{\end{figure*}}
\newenvironment{resulttablewide}{\begin{table*}[tp]\centering}{\end{table*}}

\makeatother

\hypersetup{hidelinks}
\colorlet{blue}{black}
\colorlet{red}{black}
\colorlet{magenta}{black}
\colorlet{green}{black}
\definecolor{revisionblue}{RGB}{0,70,200}
\definecolor{reviewgreen}{RGB}{0,125,45}
\definecolor{citationchange}{RGB}{200,70,0}
\definecolor{sourcechange}{RGB}{0,115,130}
\definecolor{resultheadingchange}{RGB}{115,55,155}
\definecolor{physicsclaimchange}{RGB}{0,105,75}
\definecolor{timingchange}{RGB}{0,115,145}
\definecolor{latestchange}{RGB}{170,0,100}

\newcommand{\msun}{\ensuremath{M_\odot}}

\begin{document}

\title{Fast Bayesian Updating of the Neutron-Star Equation of State with
Neural Posterior and Evidence Estimation}

\author{Prashant Thakur}
\email{prashant@yonsei.ac.kr}
\affiliation{Department of Physics, Yonsei University,
Seoul 03722, South Korea}

\date{\today}

\begin{abstract}
We present two complementary neural routes for microscopic
neutron-star EOS inference: truncated sequential neural posterior estimation
with mixture importance sampling (TSNPE+MIS), which targets one observation,
and amortized neural posterior estimation with importance sampling (A-NET+IS),
which is reused across nuclear inputs and NICER sources. Both are corrected
against the original prior and full likelihood, with the effective sample size
as a diagnostic. We test nucleonic and hyperonic relativistic mean-field models
with nuclear data, NICER, GW170817, and pQCD. For the fiducial analyses, both
agree with UltraNest in the parameter posteriors and 90\% mass--radius and
tidal-deformability bands. TSNPE+MIS band-edge differences
are only $0.024$--$0.033$~km and $1.02$--$1.03\%$, while A-NET+IS differences
are $0.019$--$0.050$~km and $0.86$--$1.13\%$. TSNPE+MIS reduces the computing time
by factors of $1.6$--$2.3$; after training, A-NET+IS is $15$--$240$ times faster
per configuration. The frozen A-NET also analyzes three NICER sources absent
from training, including the newly reported high-mass, compact
PSR~J1614$-$2230 posterior. For this source, the inferred $R_{1.4}$ agrees with
UltraNest within $0.010$~km for nucleonic matter and $0.009$~km for hyperonic
matter. We also introduce a Green-function Evidence Network (EN), a
regression-based variant of Evidence Networks and, to our knowledge, the first
Evidence Network applied to neutron-star EOS inference. After a one-time cost,
its network queries return absolute evidences in about $0.03$~s for new
nuclear data and $3.5$~s for a new NICER source; all 16 differences from UltraNest are below
$1\sigma$, and independent importance-sampling evidences agree within the quoted EN uncertainties. Under the adopted likelihood, both evidence calculations
favor nucleonic matter by Bayes factors of about $22{:}1$--$27{:}1$. Together, these results establish targeted, amortized, and
evidence-level routes for fast, likelihood-corrected EOS inference.
\end{abstract}

\maketitle

\section{Introduction}
\label{sec:intro}

Modern inference of the neutron-star equation of state (EOS)
is defined not only by the diversity of available constraints, but also by
their continual revision. Nuclear-matter information and chiral effective
field theory (chiral EFT) constrain the EOS around and below saturation density
\cite{Malik:2022zol,Thakur:2026dpr}; the Neutron Star Interior Composition
Explorer (NICER) provides source-specific
mass--radius likelihoods
\cite{Miller:2019cac,Riley:2019yda,Dittmann:2024mbo,Choudhury:2024xbk,
Mauviard:2025dmd,Salmi:2024bss,Mauviard:2026gzc}; binary neutron star (BNS)
observations probe the
tidal response \cite{LIGOScientific:2017vwq}; and perturbative quantum
chromodynamics (pQCD) restricts the admissible high-density continuation
\cite{Komoltsev:2021jzg,Gorda:2022jvk}. Because these inputs enter a joint
likelihood, adding a source, replacing a pulse-profile posterior, or changing
an adopted nuclear measurement defines a new posterior and, for model
comparison, a new Bayesian evidence. This updating burden will grow
substantially in the era of the Einstein Telescope (ET) and Cosmic Explorer
(CE), whose much larger catalogs of BNS observations will require EOS
analyses across many events and successive data releases
\cite{Punturo:2010zz,Evans:2021gyd}. Dense-matter inference is therefore
becoming a problem of repeated Bayesian updating, for which a statistically
faithful analysis must absorb evolving data without incurring the cost of a
fresh full-likelihood exploration each time.

Conventional nested sampling provides a direct
likelihood-based route to both the posterior distribution and the Bayesian
evidence and has therefore been widely used in neutron-star EOS inference
\cite{Malik:2022zol,Malik:2022jqc,Buchner:2021cql}. For a microscopic EOS
model, however, exploring the parameter space requires repeated evaluation of
the complete forward chain: the nuclear-matter properties, beta-equilibrated
EOS, stellar structure and tidal response, and the associated observational
likelihoods. This computation is tied to the particular data configuration
being analyzed. Samples from an earlier calculation can be reused only when
they retain sufficient support under the updated target; otherwise, a new
sampling run is required. As independent nuclear and astrophysical inputs
accumulate, the principal bottleneck is therefore not merely the cost of one
high-fidelity analysis, but the multiplication of that cost across many
observations and successive data configurations.

Amortization changes this scaling rather than merely
accelerating another individual inference. A conditional neural posterior
estimator can learn the mapping from changing nuclear and astrophysical inputs
to the EOS-parameter posterior once and then reuse that mapping for subsequent
observations \cite{Cranmer:2019eaq,Dax:2023ozk,Thakur:2026dpr}.
This reuse is
fundamentally different from sequential neural inference, which concentrates
simulations around one target and must repeat that procedure when the target
changes \cite{Deistler:2022tsnpe}.
Within neutron-star EOS inference, neural posterior
estimation has been applied to model-agnostic EOS representations
\cite{Carvalho:2025qie}, also for a variable number of observations, each
represented by its posterior samples \cite{Carvalho:2026nbl}, and set-based
networks have reconstructed the EOS from unordered sets of stars
\cite{Ferreira:2026bwb}. Neural likelihood estimation has enabled
inference directly from telescope spectra \cite{Brandes:2024vhw}.
The advantage of amortization therefore
comes with a central reliability requirement: every new neural posterior must
remain faithful to the exact prior and likelihood, including for observations
not encountered during training. Moreover, posterior estimation alone does
not provide the Bayesian evidence required for comparing EOS models. A
practically complete framework must consequently combine rapid reuse across
observations with per-target posterior correction and an equally efficient
route to Bayesian evidence.

A first proof of concept was established in
Ref.~\cite{Thakur:2026dpr}, where an amortized neural posterior estimator
reproduced PyMultiNest posteriors for the microscopic couplings of
two relativistic mean-field (RMF) models: the density-dependent DDB and
nonlinear RMF-NL parametrizations, constrained by nuclear-saturation
properties, chiral-EFT
pure-neutron-matter pressures, and a maximum-mass bound.
That study, however, conditioned only on compact
summary constraints and did not include full source-dependent NICER
mass--radius likelihoods, gravitational-wave (GW) tidal information, or pQCD
constraints. It also left open two requirements identified above: correcting
each neural posterior against the original likelihood and obtaining Bayesian
evidences without repeating conventional sampling. Whether a single frozen
estimator can analyze genuinely unseen stellar observations, retain the
accuracy of conventional Bayesian inference, and simultaneously support EOS
model comparison has therefore remained unresolved for microscopic EOS
inference.

In this work, we develop two
complementary, likelihood-corrected neural routes for microscopic neutron-star
EOS inference. Truncated sequential neural posterior estimation with mixture
importance sampling (TSNPE+MIS) concentrates new simulations around a
specified observation, providing a simulation-efficient posterior for a
focused analysis \cite{Deistler:2022tsnpe}. Amortized neural posterior
estimation with importance sampling (A-NET+IS) instead uses flow-matching
posterior estimation to learn a reusable EOS-parameter proposal conditioned
jointly on nuclear measurements and sets of NICER
mass--radius posterior samples
\cite{Dax:2023ozk}. Within a given EOS family,
frozen A-NET proposals are applied without retraining to shifted nuclear
inputs and to three NICER sources that were absent from training, including
the newly reported high-mass, compact PSR~J1614$-$2230 posterior
\cite{Mauviard:2026gzc}. For both neural routes, samples are reweighted using
the exact prior, normalized proposal density, and complete multimessenger
likelihood, while the effective sample size provides a target-specific
diagnostic of the correction \cite{Dax:2022pxd}. We compare both neural posteriors
one-to-one with UltraNest \cite{Buchner:2021cql}, forming three algorithmically
distinct inference routes with none treated as ground truth.
In parallel, a Green-function Evidence Network (EN),
constructed in the spirit of Evidence Networks \cite{Jeffrey:2023stk},
provides absolute Bayesian evidences across configurations with changed
nuclear data or NICER sources. It replaces the classifier of the original
construction by regression on importance-sampling evidence labels; to the best of our
knowledge, this is the first application of the Evidence-Network approach to
neutron-star EOS inference.
We first establish the
complete framework for nucleonic DDB matter and then repeat the analysis for
the hyperonic DDB$\Lambda\Xi^-$ extension. The latter introduces additional
microscopic degrees of freedom and a different high-density particle
composition, providing a controlled demonstration that the framework is not
restricted to purely nucleonic matter.

The paper is organized as follows.
Section~\ref{sec:method} presents the methodology, including the EOS models,
multimessenger likelihood, and the UltraNest, A-NET, TSNPE, and Evidence
Network inference procedures. Section~\ref{sec:results} presents the results
for nucleonic and hyperonic matter, including the inferred parameter and
stellar-observable posteriors, Bayesian evidences, tests with NICER sources
absent from training, and computational performance.
Section~\ref{sec:conclusions} presents the conclusions and outlook.
Appendix~\ref{app:hyperonic-nuclear-amortization} presents the nuclear-data
amortization for hyperonic matter, Appendix~\ref{app:anet-calibration} the
calibration of the A-NET proposals, and Appendix~\ref{app:numerics} the
numerical convergence and robustness tests.

\section{Methodology}
\label{sec:method}

\begin{figure*}[tp]
  \centering
  \includegraphics[width=\textwidth]{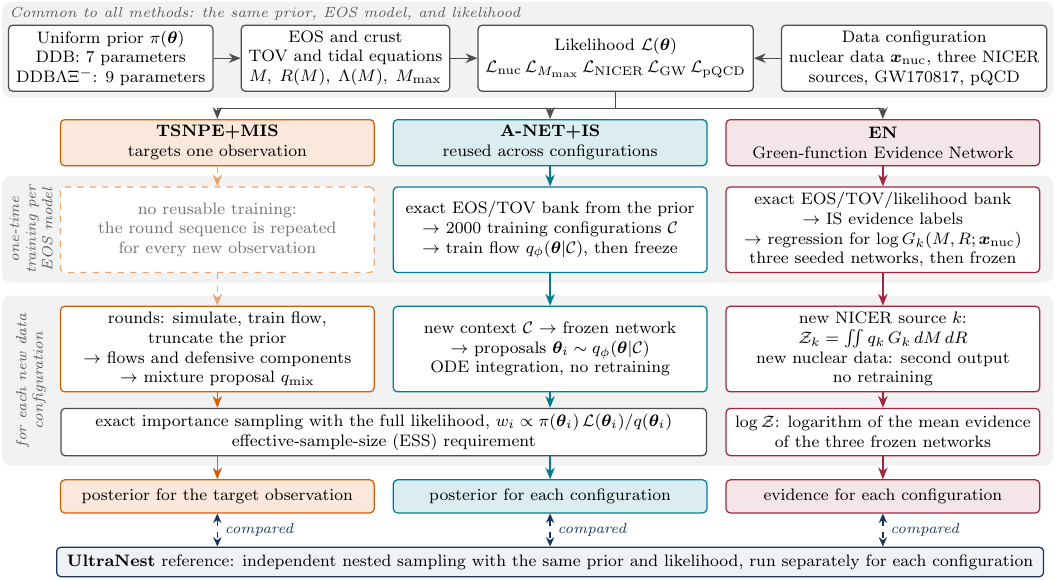}
  \caption{Workflow of TSNPE+MIS, A-NET+IS, and the Green-function Evidence
  Network (EN). All methods use the same prior, EOS model, and likelihood;
  UltraNest provides the independent reference.}
  \label{fig:workflow}
\end{figure*}

Figure~\ref{fig:workflow} summarizes the inference workflow. All methods use
the prior, EOS model, and likelihood of Secs.~\ref{sec:method:eos}
and~\ref{sec:method:like}, and their results are compared with independent
UltraNest calculations (Sec.~\ref{sec:method:ultranest}). The methods differ
in which steps are carried out once for each EOS model and which are repeated
for each new data configuration, that is, for shifted nuclear data or a
substituted NICER source. TSNPE (Sec.~\ref{sec:method:tsnpe}) has no reusable
training stage: its simulation and training rounds are repeated for every new
observation. A-NET (Sec.~\ref{sec:method:anet}) is trained once, and a new
configuration requires only a query of the frozen network to obtain an
importance-sampling proposal. Both posterior routes end with importance
sampling against the full likelihood and are checked with the effective sample
size. The EN (Sec.~\ref{sec:method:en}) is also trained once; for a new
configuration, the three frozen networks give the evidence without retraining,
either by integrating the density of a substituted NICER source against the
learned Green-function kernel or through a second network output for shifted
nuclear data.

\subsection{Equation of state}
\label{sec:method:eos}

We use the nucleonic DDB EOS of
Ref.~\cite{Malik:2022zol}, described by a seven-parameter vector $\bm\theta$
with a uniform prior. The vector comprises six parameters of the
density-dependent meson--nucleon coupling functions and the saturation density
$\rho_0$, which sets their reference density. The six coupling-function
parameters and their prior ranges follow our companion
study~\cite{Thakur:2026dpr}, in which the DDB inference was validated against
nested sampling. The core is joined to a Baym--Pethick--Sutherland (BPS) outer
crust~\cite{Baym:1971pw} through a polytropic inner-crust interpolation, and the
Tolman--Oppenheimer--Volkoff and tidal equations are integrated along the stable
branch $\{M,R(M),\Lambda(M)\}$. We additionally use the hyperonic extension
DDB$\Lambda\Xi^-$ of
Ref.~\cite{Malik:2022jqc}, which augments the same
nucleonic core with $\Lambda$ and $\Xi^-$ hyperons, with vector couplings
fixed by SU(6) symmetry and scalar coupling ratios calibrated to hypernuclear
data~\cite{Fortin:2017cvt,Fortin:2020qin}.
The prior ranges adopted for both EOS models are
summarized in Table~\ref{tab:eos-priors}.

\begin{table*}[t]
  \centering
  \caption{Uniform prior ranges used in this work for the nucleonic DDB and
  hyperonic DDB$\Lambda\Xi^-$ EOS parameters. The seven nucleonic parameters
  are common to both models; the two scalar hyperon-to-nucleon coupling ratios
  enter only the hyperonic model. The saturation density $\rho_0$ is given in
  ${\rm fm}^{-3}$; the meson--nucleon couplings $\Gamma_{M,0}$,
  density-dependence parameters $a_M$, and coupling ratios $x_{\sigma Y}$ are
  dimensionless.}
  \label{tab:eos-priors}
  \footnotesize
  \setlength{\tabcolsep}{3.5pt}
  \renewcommand{\arraystretch}{1.10}
  \begin{tabular*}{\textwidth}{@{\extracolsep{\fill}}lcccccccc@{}}
    \toprule
    Model & Parameter & Prior range & Parameter & Prior range &
    Parameter & Prior range & Parameter & Prior range \\
    \midrule
    Nucleonic DDB
      & $a_\sigma$ & $[0,\,0.300]$
      & $a_\omega$ & $[0,\,0.298446]$
      & $a_\rho$ & $[0,\,1.300]$
      & $\rho_0\;({\rm fm}^{-3})$ & $[0.140,\,0.170]$ \\
    {}
      & $\Gamma_{\sigma,0}$ & $[6.500,\,13.4895]$
      & $\Gamma_{\omega,0}$ & $[7.500,\,14.500]$
      & $\Gamma_{\rho,0}$ & $[5.000,\,12.5471]$
      & \multicolumn{2}{c}{} \\
    Additional DDB$\Lambda\Xi^-$
      & $x_{\sigma\Lambda}$ & $[0.609,\,0.622]$
      & $x_{\sigma\Xi}$ & $[0.309,\,0.322]$
      & \multicolumn{4}{c}{} \\
    \bottomrule
  \end{tabular*}
\end{table*}

\subsection{Likelihood and data}
\label{sec:method:like}

The likelihood factorizes into a nuclear-matter term and an astrophysical
term, each depending only on the EOS outputs of
Sec.~\ref{sec:method:eos}: the seven nuclear-matter observables
$g(\bm\theta)$, and the stable mass--radius--tidal branch up to the
maximum mass $M_{\max}(\bm\theta)$,
\begin{align}
\mathcal{L}(\bm\theta) &= \mathcal{L}_{\rm nuc}[g(\bm\theta)]\,
\mathcal{L}_{\rm astro}(\bm\theta), \label{eq:likelihood}\\
\mathcal{L}_{\rm astro} &\equiv \mathcal{L}_{M_{\max}}\,
\mathcal{L}_{\rm NICER}\,\mathcal{L}_{\rm GW}\,\mathcal{L}_{\rm pQCD}.
\end{align}

\emph{Nuclear-matter kernel.} For the evidence convention used
throughout this work, the nuclear-matter likelihood is the unnormalized
diagonal-Gaussian kernel
\[
\mathcal{L}_{\rm nuc}[g(\bm\theta);\bm{x}_{\rm obs}]
=\exp\!\left[-\frac{1}{2}
\sum_{i=1}^{7}
\left(\frac{g_i(\bm\theta)-x_{{\rm obs},i}}{\sigma_i}\right)^2\right].
\]
The parameter-independent Gaussian normalization is omitted consistently
by all inference methods. The kernel constrains the
saturation density, binding energy, incompressibility, symmetry energy, and the
chiral-EFT pure-neutron-matter pressure at $0.08$, $0.12$, and
$0.16\,{\rm fm}^{-3}$, with the same central values and widths as
\cite{Thakur:2026dpr}. The four nuclear-matter shifts of
Sec.~\ref{sec:results} move one component of $\bm{x}_{\rm obs}$ at a time,
with $\Sigma$ fixed.

\emph{Maximum-mass constraint.} $\mathcal{L}_{M_{\max}}$ is a soft logistic floor
centered at $2.0\,\msun$ with width $0.05\,\msun$, requiring support for
the observed population of ${\sim}2\,\msun$ pulsars
\cite{Antoniadis:2013pzd,Fonseca:2021wxt}, without
imposing a hard cutoff. Because the mass of PSR~J0740$+$6620 also enters its NICER
posterior, this factor counts part of that information twice; removing it
lowers the median $M_{\max}$ by at most $0.04\,\msun$ and the median $R_{1.4}$
by at most $0.06$~km (Appendix~\ref{app:numerics}).

\emph{NICER.} We construct $\mathcal{L}_{\rm NICER}$ from Gaussian kernel
density estimates $\hat p_s(M,R)$ of the published mass--radius posteriors, as is
common in neutron-star EOS inference \cite{Cartaxo:2025jpi,Huang:2024rfg}. The
fiducial configuration includes PSR~J0030$+$0451 \cite{Miller:2019cac},
PSR~J0740$+$6620 \cite{Dittmann:2024mbo}, and PSR~J0437$-$4715
\cite{Choudhury:2024xbk}; Table~\ref{tab:inputs} lists the data products and
the priors of their analyses, which we keep. For each source, the likelihood is
the average of $\hat p_s$ along the stable branch, that is, with a uniform
stellar-mass prior on $[M_{\rm lo},M_{\max}(\bm\theta)]$,
\begin{equation}
\mathcal L_{{\rm NICER},s}(\bm\theta)=\frac{1}{M_{\max}-M_{\rm lo}}
\int_{M_{\rm lo}}^{M_{\max}}\hat p_s\bigl(M,R(M;\bm\theta)\bigr)\,dM,
\label{eq:nicer-like}
\end{equation}
with $M_{\rm lo}=1\,\msun$, and $\mathcal{L}_{\rm NICER}$ is the product over
the three sources. All compared methods use this same likelihood;
Appendix~\ref{app:numerics} gives its numerical settings and tests them,
together with the kept priors.

\emph{GW170817.} $\mathcal{L}_{\rm GW}$ uses a four-dimensional Gaussian kernel
density estimate $\hat p_{\rm GW}$ of the GWTC-1 low-spin posterior samples
\cite{LIGOScientific:2018mvr,LIGOScientific:2019gwtc1data} on source-frame
chirp mass $\mathcal M$, mass ratio $q$, and the tidal deformabilities
$(\Lambda_1,\Lambda_2)$. With the chirp mass fixed at its marginal mode,
$\mathcal M_\star\approx1.186\,\msun$,
\begin{equation}
\mathcal L_{\rm GW}(\bm\theta)=\int_{0.7}^{1}\hat p_{\rm GW}\bigl(\mathcal
M_\star,q,\Lambda(m_1;\bm\theta),\Lambda(m_2;\bm\theta)\bigr)\,dq,
\label{eq:gw-like}
\end{equation}
where $m_{1,2}$ follow from $(\mathcal M_\star,q)$. Fixing the chirp mass and
keeping the prior of the original analysis each change the median $R_{1.4}$ by
at most $0.06$~km (Appendix~\ref{app:numerics}).

\emph{pQCD.} $\mathcal{L}_{\rm pQCD}$ is the causal-connection likelihood
to the pQCD EOS
\cite{Komoltsev:2021jzg,Gorda:2022jvk}, evaluated at
$1.2\,{\rm fm}^{-3}$ \cite{Cartaxo:2025jpi} with the implementation of
\cite{Huang:2024rfg}: the log-fraction of a fixed $400$-point
renormalization-scale grid admitting a stable, causal connection.

\emph{Substituted sources.} PSR~J0614$-$3329 \cite{Mauviard:2025dmd}
replaces PSR~J0437$-$4715, PSR~J1231$-$1411 \cite{Salmi:2024bss}, analyzed
with a uniform $10$--$14$~km radius prior, replaces PSR~J0030$+$0451, and
PSR~J1614$-$2230
\cite{Mauviard:2026gzc} replaces PSR~J0740$+$6620. Each substituted
source enters $\mathcal{L}_{\rm NICER}$ through its own kernel density
estimate, built identically to the fiducial sources; the other two NICER
sources and all remaining likelihood terms are unchanged in each case.

\begin{table*}[t]
  \centering
  \caption{Likelihood inputs. Upper block: central values and $1\sigma$
  widths of the seven nuclear-matter constraints in the A1 configuration
  \cite{Thakur:2026dpr}; $P_{\rm PNM}$ is the pure-neutron-matter pressure at
  the indicated density. Lower block: astrophysical data products and the
  priors of their analyses; ``radio'' denotes a Gaussian mass prior from radio
  timing.}
  \label{tab:inputs}
  \scriptsize
  \setlength{\tabcolsep}{2pt}
  \renewcommand{\arraystretch}{1.12}
  \begin{tabular*}{\textwidth}{@{\extracolsep{\fill}}lllll@{}}
    \toprule
    \multicolumn{5}{@{}l}{\textit{Nuclear-matter constraints}} \\
    $\rho_0$ (fm$^{-3}$) & $E_0$ (MeV) & $K_0$ (MeV) & $J_{\rm sym}$ (MeV) &
    $P_{\rm PNM}$ at $0.08$, $0.12$, $0.16$~fm$^{-3}$ (MeV\,fm$^{-3}$) \\
    $0.153\pm0.005$ & $-16.1\pm0.2$ & $230\pm40$ & $32.5\pm1.8$ &
    $0.51\pm0.19$, \ $1.24\pm0.61$, \ $2.49\pm1.38$ \\
    \midrule
    Data product & Analysis & Mass prior & Radius prior & Samples \\
    \midrule
    PSR~J0030$+$0451 & \cite{Miller:2019cac}, two spots & flat, $1.0$--$2.4\,\msun$
      & flat in $GM/c^2R\in[0.125,0.3125]$ & $4.0\times10^5$ \\
    PSR~J0740$+$6620 & \cite{Dittmann:2024mbo}, NICER$+$XMM & radio, $2.08\pm0.09\,\msun$
      & flat in $c^2R/GM\in[3.2,8.0]$ & $9.8\times10^5$, weighted \\
    PSR~J0437$-$4715 & \cite{Choudhury:2024xbk}, CST$+$PDT & radio, $1.418\pm0.044\,\msun$
      & flat in $R$ & $2.2\times10^5$ \\
    PSR~J0614$-$3329 & \cite{Mauviard:2025dmd}, ST$+$PDT & radio, $1.44\pm0.07\,\msun$
      & flat, $6$--$16$~km & $2.0\times10^5$ \\
    PSR~J1231$-$1411 & \cite{Salmi:2024bss} & radio, broad & flat, $10$--$14$~km
      & $1.1\times10^6$, weighted \\
    PSR~J1614$-$2230 & \cite{Mauviard:2026gzc}, ST-U & radio, $1.937\pm0.014\,\msun$
      & flat, $6$--$16$~km & $5.0\times10^5$ \\
    GW170817 & \cite{LIGOScientific:2019gwtc1data}, low spin & flat in $m_{1,2}$
      & flat in $\Lambda_{1,2}\in[0,5000]$ & $8.1\times10^3$ \\
    \bottomrule
  \end{tabular*}
\end{table*}

\subsection{UltraNest}
\label{sec:method:ultranest}
UltraNest is an established reactive nested-sampling framework
for posterior inference and evidence
calculation~\cite{Skilling:2006gxv,Buchner:2021cql};
applications involving neutron stars include NICER
pulse-profile inference for PSR~J0740$+$6620~\cite{Hoogkamer:2025ype} and
gravitational-wave parameter inference for the binary-neutron-star merger
GW170817~\cite{Buchner:2024rjd}. We use
UltraNest~4.5.0 with the vectorized \texttt{ReactiveNestedSampler}, a minimum of
$1000$ live points, $30$ bootstrap integrators, and resumable HDF5 output. The
default convergence targets are retained: $\Delta\ln Z=0.5$, a
Kullback--Leibler divergence of $d_{\rm KL}=0.5$, a remaining-evidence
fraction of $0.01$, and a minimum posterior
effective sample size (ESS) of $400$.
UltraNest, A-NET+IS, and TSNPE+MIS use the same uniform
prior (Table~\ref{tab:eos-priors}) and the same likelihood,
Eq.~\eqref{eq:likelihood}, including every constraint described above. In $24$
of the $27$ UltraNest runs, the sampled likelihood also applied the sample weights
of PSR~J0740$+$6620 (and, in the J1231 configuration, of PSR~J1231$-$1411) inside
the kernel density estimate, so that these weights entered twice; each such run
was reweighted to Eq.~\eqref{eq:likelihood} with the exact likelihood ratio,
which raises $\ln Z$ by $0.15$--$0.23$ with a Monte Carlo error below $0.002$
that is included in the quoted uncertainties. For each
shifted or substituted configuration, all methods evaluated use the same
correspondingly modified likelihood; their posteriors and evidences are
therefore compared on an equal footing.

\subsection{Amortized neural posterior estimation (A-NET)}
\label{sec:method:anet}
For the nucleonic DDB model, let $\bm\theta$ denote
the seven-dimensional parameter vector. For a given
observational configuration, we collect the conditioning information into
\begin{align}
\mathcal C &\equiv
 \left\{
 \underbrace{\bm x_{\rm nuc}}_{\text{nuclear}},
 \underbrace{\mathcal X_1,\mathcal X_2,\mathcal X_3}_{\text{NICER}},
 \underbrace{g_\Lambda}_{\text{GW}},
 \underbrace{\rho_{\rm pQCD}}_{\text{pQCD}}
 \right\},
 \label{eq:anet-context}\\
\mathcal X_s &\equiv \{(M_{sk},R_{sk})\}_{k=1}^{256},
\end{align}
where $\mathcal X_s$ is an unordered set of
mass--radius posterior samples for NICER source $s$ and $\bm x_{\rm nuc}$ contains the seven nuclear-matter constraints.
The control $g_\Lambda$ rescales the tidal
deformabilities entering the GW likelihood as
$\Lambda\rightarrow\Lambda e^{-g_\Lambda}$, while $\rho_{\rm pQCD}$ specifies
the density at which the pQCD likelihood is evaluated. For every $\mathcal C$,
the target is the same posterior sampled by the conventional Bayesian methods,
\begin{equation}
\begin{aligned}
p(\bm\theta\mid\mathcal C)
={}&\frac{\pi(\bm\theta)}{\mathcal Z(\mathcal C)}
 \mathcal L_{\rm nuc}(\bm\theta;\bm x_{\rm nuc})
 \mathcal L_{M_{\max}}(\bm\theta)\\
&\times \mathcal L_{\rm GW}(\bm\theta;g_\Lambda)
 \mathcal L_{\rm pQCD}(\bm\theta;\rho_{\rm pQCD})\\
&\times\prod_{s=1}^{3}
 \mathcal L_{{\rm NICER},s}(\bm\theta;\mathcal X_s).
\end{aligned}
\label{eq:anet-target}
\end{equation}
A-NET learns a conditional approximation
$q_\phi(\bm\theta\mid\mathcal C)$ over a family of such configurations.
Changing $\bm x_{\rm nuc}$ specifies a different set of nuclear constraints,
whereas changing any $\mathcal X_s$ specifies a different NICER configuration.
The two additional controls were also varied during
training. Every result reported here uses the fiducial values $g_\Lambda=0$
and $\rho_{\rm pQCD}=1.2\,{\rm fm}^{-3}$, while the maximum-mass factor remains
fixed. These inputs can be passed to the same network without retraining.
Importantly, $q_\phi$ is used as an amortized proposal; the reported posterior
is subsequently corrected using the complete exact likelihood, including the
GW and pQCD factors at the reported values, and the
proposal density.

The training data are generated once, before any
configuration is analyzed, and are reused for all configurations. We start from
a bank of $N_0=660\,000$ samples drawn from the uniform DDB prior. For every
sample we precompute the seven nuclear observables,
the stable mass--radius--tidal sequence $\{M,R(M),\Lambda(M)\}$, $M_{\max}$,
and the likelihood ingredients required by Eq.~\eqref{eq:likelihood}. Because
the posterior occupies a small fraction of the prior volume, we supplement the
original bank with $254\,966$ samples from three predefined enrichment regions:
one concentrated near the measured nuclear constraints, with all seven nuclear
predictions within $3\sigma$ of the observation (screened from
$1.6\times10^{8}$ prior draws), and two covering the low- and high-$R_{1.4}$
tails, $R_{1.4}\in[9.4,11.2]$~km or $[14.8,18.0]$~km with
$M_{\max}\geq1.9\,\msun$ (preselected with a linear predictor of $R_{1.4}$ from
$2.64\times10^{6}$ and $1.32\times10^{6}$ prior draws). This gives a total of $914\,966$ parameter
points. Since each region is obtained by screening a known number $M_s$ of
independent prior draws, its sampling intensity and correction are known
exactly:
\begin{align}
\lambda(\bm\theta)&\propto\pi(\bm\theta)
 \left[N_0+\sum_s M_s\mathbf 1_{S_s}(\bm\theta)\right],
 \label{eq:anet-bank-intensity}\\
r(\bm\theta)&=
 \frac{N_0}{N_0+\sum_sM_s\mathbf 1_{S_s}(\bm\theta)}.
 \label{eq:anet-bank-correction}
\end{align}
The factor $r(\bm\theta)$ is included in every subsequent bank weight, so the
enrichment improves sampling density without changing the uniform-prior target.
For the observed nuclear-matter kernel, this procedure increases the
conditional ESS from $10.1$ to $4384$.

We construct $2000$ training configurations from the three NICER mass--radius
sample sets. The first $200$ reproduce A1 with
$g_\Lambda=0$ and $\rho_{\rm pQCD}=1.2\,{\rm fm}^{-3}$; in the remaining
$1800$, the two controls are drawn uniformly from $g_\Lambda\in[-0.58,0.58]$
and $\rho_{\rm pQCD}\in[1.02,1.38]\,{\rm fm}^{-3}$, and none, one, or all three
sample sets are
transformed in a $25{:}50{:}25$ ratio. Each transformation
shifts and rescales a sample set about its weighted centre,
\begin{align}
M'&=\bar M_s+a_M(M-\bar M_s)+\Delta M,\\
R'&=\bar R_s+a_R(R-\bar R_s)+\Delta R,
\label{eq:anet-cloud-transform}
\end{align}
with $\Delta M\in[-0.5,0.25]~\msun$,
$\Delta R\in[-1.6,1.6]~{\rm km}$, and $a_M,a_R\in[0.28,2.2]$.
The GW, pQCD, and NICER factors therefore follow the
training configuration, while the maximum-mass factor remains fixed. For each
configuration $j$, the bank points are weighted as
\begin{equation}
\begin{aligned}
\bar w_i^{(j)}\propto
{}&r(\bm\theta_i)\,
\mathcal L_{M_{\max}}(\bm\theta_i)\\
&\times\mathcal L_{\rm GW}(\bm\theta_i;g_\Lambda^{(j)})\,
\mathcal L_{\rm pQCD}(\bm\theta_i;\rho_{\rm pQCD}^{(j)})\\
&\times\prod_{s=1}^{3}
\mathcal L_{{\rm NICER},s}
(\bm\theta_i;\mathcal X_s^{(j)}),
\qquad \sum_i\bar w_i^{(j)}=1.
\end{aligned}
\label{eq:anet-scenario-weight}
\end{equation}
Configurations with $N_{\rm eff}<250$ are discarded: $1937$ of the $3937$ drawn
configurations, including every draw with $g_\Lambda>0.022$, so that the
training set covers $g_\Lambda\in[-0.58,0.022]$; all benchmarks of
Sec.~\ref{sec:results} use $g_\Lambda=0$ and $\rho_{\rm pQCD}=1.2\,{\rm fm}^{-3}$.
We then select bank
points according to
\begin{align}
\Pr\!\left(\bm\theta_{jn}=\bm\theta_i\mid\mathcal C_j\right)
 &=\bar w_i^{(j)},\\
\bm x_{{\rm nuc},jn}&=g(\bm\theta_{jn})+\bm\epsilon_{jn},
\qquad \bm\epsilon_{jn}\sim\mathcal N(\bm0,\Sigma).
\label{eq:anet-training-pairs}
\end{align}
The resulting dataset therefore varies both the NICER inputs and the nuclear
constraints, and contains $9\,008\,849$ training and $452\,721$ validation
pairs.

Each NICER sample set is encoded by a shared, permutation-invariant Deep Sets
module~\cite{Zaheer:2017wmg}. A three-layer pointwise network of width $128$ acts on every $(M,R)$
sample together with its source label; averaging the resulting $256$ point
embeddings removes any dependence on their ordering. A second map compresses
the average to a $64$-component source representation. The three source
representations are concatenated with the standardized seven-component nuclear
vector and the two standardized GW and pQCD controls
to form the $201$-component context $h(\mathcal C)$. This context and a
$16$-component sinusoidal time embedding condition a vector field with four hidden layers of
width $512$, which transports a seven-dimensional standard normal distribution
to the conditional posterior $q_\phi(\bm\theta\mid\mathcal C)$.

We implement the posterior-estimation component of A-NET using flow-matching
posterior estimation (FMPE)~\cite{Dax:2023ozk}, which applies conditional flow
matching~\cite{Lipman:2023hef} to learn
$q_\phi(\bm\theta\mid\mathcal C)$. Let $\widetilde{\bm\theta}$ be a
standardized training sample,
$\bm z\sim\mathcal N(\bm0,I)$, and $t\sim\mathcal U(0,1)$. We define
\begin{equation}
\bm y_t=\bigl[1-(1-\sigma_{\min})t\bigr]\bm z
       +t\widetilde{\bm\theta},\qquad
\bm u_t=\widetilde{\bm\theta}-(1-\sigma_{\min})\bm z,
\label{eq:anet-flow-path}
\end{equation}
and minimize
\begin{equation}
\mathcal L_{\rm FM}=
\mathbb E\!\left[
\left\|v_\phi(\bm y_t,t,h(\mathcal C))-\bm u_t\right\|_2^2
\right].
\label{eq:anet-flow-loss}
\end{equation}
We use $\sigma_{\min}=10^{-3}$ and train for $600$ epochs with
Adam~\cite{Kingma:2014vow} and a
cosine-annealed learning rate initialized at
$10^{-3}$~\cite{Loshchilov:2016tgk}, redrawing the nuclear noise
at every epoch. Posterior samples are obtained by integrating
$d\bm y_t/dt=v_\phi$ from $t=0$ to $1$ with a $64$-step Heun
scheme~\cite{Hairer:1993ode} and then
undoing the parameter standardization. The density accumulated along the same
$64$-step trajectory differs from the exact density of the discrete map mainly
by a constant offset, which cancels on normalization; the remaining density
dependence changes the tested posterior summaries by less than their bootstrap
uncertainties, and the corrected posterior agrees with $256$- and $1024$-step
integration within its Monte Carlo errors (Appendix~\ref{app:numerics}). A new nuclear--NICER context therefore
requires only a new integration of the ordinary differential equation (ODE),
not retraining.

For the hyperonic DDB$\Lambda\Xi^-$ model, we use the
same Deep Sets encoder architecture for three $256$-point NICER sample sets and the
same vector-field architecture, with four hidden layers of width $512$. The
common scenario set comprises $2000$ configurations constructed solely from
the three A1 NICER sources, partitioned into $1900$ training and $100$
validation configurations. Their posterior rows are drawn, with exact weights,
from $6\times10^{5}$ uniform prior draws ($415{,}681$ with a valid stellar
branch), $83{,}764$ draws with all seven nuclear predictions within $3\sigma$
of the observation (screened from $1.6\times10^{8}$ prior draws), and
$3\times10^{5}$ draws from the defensive density $q_{\rm def}$ defined below,
combined as a deterministic mixture with exactly known sampling intensities. The
$199$-component context of the network contains the three
encoded sample sets and the seven nuclear inputs, with no GW or pQCD controls. We
train two heads in bounded coordinates: a nine-dimensional head for all
physical parameters and a seven-dimensional head for the DDB parameters, with
the two hyperon-coupling ratios retaining their exact uniform densities. Both
head densities include the corresponding coordinate-transformation
Jacobians. The normalized proposal is
$q_{\rm hyp}=0.35\,q_9+0.55\,q_7+0.10\,q_{\rm def}$, where
$q_{\rm def}=0.90\,q_{t,48}+0.10\,\pi$ combines a $48$-component multivariate
Student-$t$ mixture, fitted to an exact-likelihood exploration of the parameter
space, with the uniform prior $\pi$. Every draw is assigned an importance
weight using the complete mixture density, and both sampling from the flow
heads and evaluating their densities use $1024$-step Heun integration.

To remove any residual approximation error from the amortized posterior, we use
A-NET as a neural IS proposal~\cite{Dax:2022pxd}. For a given
context $\mathcal C$, we draw
$\bm\theta_i\sim q_\phi(\bm\theta\mid\mathcal C)$ and evaluate each sample using
the same prior and full likelihood employed by UltraNest. The unnormalized and
normalized importance weights are
\begin{equation}
\widetilde w_i=
\frac{\pi(\bm\theta_i)\,
\mathcal L_{\rm full}(\bm\theta_i;\mathcal C)}
{q_\phi(\bm\theta_i\mid\mathcal C)},
\qquad
w_i=\frac{\widetilde w_i}{\sum_j\widetilde w_j},
\label{eq:anet-is-weights}
\end{equation}
where $\mathcal L_{\rm full}$ denotes the complete likelihood in
Eq.~\eqref{eq:anet-target}. Because the adopted prior is uniform within the
parameter bounds, it contributes a constant for in-bound samples and zero
weight outside them. Out-of-bound draws are kept in all sums with zero weight,
so the proposal density keeps its normalization over all draws. For each flow component, the
proposal density is obtained by integrating its learned ODE together with the
instantaneous change-of-variables relation~\cite{Chen:2018wjc},
\begin{equation}
\frac{d}{dt}\log q_t(\bm y_t\mid\mathcal C)
=-\nabla_{\bm y}\!\cdot
v_\phi(\bm y_t,t,h(\mathcal C)).
\label{eq:anet-log-density}
\end{equation}
For each flow head, we accumulate the exact divergence
over all learned coordinates rather than using a stochastic trace estimate.
The relevant coordinate-transformation Jacobians are included in the
physical-space head densities, which are combined with any analytic components
to form the complete proposal density. Weight concentration is quantified by the
ESS,
\begin{equation}
N_{\rm eff}
=\frac{\left(\sum_i\widetilde w_i\right)^2}
{\sum_i\widetilde w_i^2}
=\frac{1}{\sum_i w_i^2}.
\label{eq:anet-ess}
\end{equation}
Each nucleonic run evaluates $8000$ A-NET proposals
and requires $N_{\rm eff}/N\geq0.005$; all reported nucleonic results reach
$N_{\rm eff}\approx400$--$3300$ (Sec.~\ref{sec:results:computational-cost}). For $K_0=260$~MeV,
$J_{\rm sym}=36$~MeV, and PSR~J1614$-$2230, four independently seeded runs are
pooled, yielding $32{,}000$ proposals. The hyperonic analyses use $25{,}000$
proposals per configuration, increased to $50{,}000$ for
$J_{\rm sym}=29$~MeV and PSR~J1231$-$1411, and require
$N_{\rm eff}\geq1000$ together with $\max_i w_i\leq0.01$ for the complete
sample. Failure of the applicable gate triggers an out-of-coverage warning, and no
A-NET result is reported for that
configuration. Its posterior must instead be obtained with a full sampler;
extending A-NET to such configurations requires augmenting the training
distribution and retraining the network. For every accepted run, the
normalized weights define the final A-NET+IS posterior. Equal-weight samples
used to calculate credible intervals and derived stellar observables are
obtained by resampling the A-NET proposals according to $w_i$; the
unweighted proposal draws are not used for scientific
results. For the fiducial nucleonic configuration, the
mass--tidal-deformability band and corresponding $\Lambda$ summaries are
obtained from a larger A-NET-seeded importance-sampling calculation: a Gaussian
mixture fitted to $7.7\times10^{5}$ A-NET+IS draws from several runs for this
configuration (ESS $3.4\times10^{4}$) supplies $65\%$ of the proposal,
a second Gaussian mixture fitted to their low-$\Lambda$ tail supplies $30\%$,
and the uniform prior supplies $5\%$. The resulting $320{,}000$ fresh draws
are weighted using the exact target and complete mixture density and must
satisfy a total ESS of at least $15{,}000$ and a minimum low-$\Lambda$-tail
ESS of $2000$; the A-NET+IS samples define only the proposal and do not enter
the resulting estimate.

\subsection{Truncated sequential neural posterior estimation
(TSNPE)}
\label{sec:method:tsnpe}

Alongside UltraNest and A-NET+IS,
TSNPE~\cite{Deistler:2022tsnpe} provides a third route to the EOS-parameter
posterior. All three reported posteriors use the same prior and full
multimessenger likelihood. Unlike A-NET, TSNPE is sequential rather than
amortized: each round's simulations and training target one specific
observation, so a new observation requires repeating the round sequence
rather than evaluating a fixed network.

A neural spline flow~\cite{Durkan:2019nsq}
($10$ transforms, $256$ hidden units per layer) is
trained round by round using neural posterior
estimation. The round-$0$ proposal is
constructed separately for the two compositions. For
nucleonic DDB, an auxiliary nuclear-plus-$M_{\max}$ flow trained on
prior-generated rows defines the first restricted prior.
For hyperonic DDB$\Lambda\Xi^{-}$, the round-$0$
simulations are drawn from a TSNPE-owned defensive proposal comprising
Gaussian mixtures and a $10\%$ uniform-prior component. The proposal is fitted
to prior-generated EOS/TOV rows under the astrophysical likelihood alone, and
$3\times10^5$ candidates yield $947$ accepted simulations. These restrictions define only the simulation
proposals, not the reported posteriors. For the
nucleonic DDB analysis, round $0$ uses $3\times10^4$ simulations. Each
subsequent round uses $2\times10^4$ simulations and truncates the prior to the
$10^{-4}$-quantile high-posterior-density region of the preceding round's
posterior, with at most $6$ rounds in total. From round $2$ onward, the
iteration stops when every DDB median shift and $90\%$ width change is below
$5\%$ of the preceding $90\%$ width. For the
hyperonic analysis, two truncated rounds follow. The first and second rounds
use, respectively, the $10^{-4}$ and $10^{-2}$ density quantiles of the
preceding estimator, and each evaluates $2\times10^4$ candidates using the same
neural-spline-flow architecture. Two independently seeded flows are then
trained on all $1069$ accepted simulations. The full-support MIS step below
restores the original full-prior target.

The trained flow is never reported as a posterior on its own: every
TSNPE result quoted in this work is corrected by exact IS
against the same likelihood used by the classical samplers,
with the normalized neural-spline-flow proposal density
evaluated directly through its invertible transformations. When a single
flow's importance weights are too
concentrated for a reliable estimate -- diagnosed by its own
ESS -- we fall back to a defensive balance-heuristic mixture.
For nucleonic A1, the final defensive mixture uses
$60{,}000$ trained-flow draws, $15{,}000$ draws from a prior-truncated
Gaussian with the initial flow-IS posterior standard deviations inflated by
$2.5$, and $30{,}000$ draws from a $1.5$-inflated prior-truncated Gaussian
fitted to the intermediate two-component posterior.
For hyperonic A1, the two flows provide the starting
point for a sequence of exact-likelihood MIS stages. The final proposal,
$q_{\rm mix}$, contains $1.5\times10^5$ draws. Of these, $90\%$ come from a
cross-validated defensive proposal consisting of a four-component Gaussian
mixture in bounded-logit coordinates, a Student-$t$ component, and a
uniform-prior component. The remaining $10\%$ come from a parent proposal that
combines two adaptively trained flow ensembles, two Student-$t$ components,
and the uniform prior. All fitted components use exclusively
importance-weighted exact rows generated in preceding TSNPE stages; no A-NET,
EN, or UltraNest product enters the procedure. The resulting
sample has $N_{\rm eff}=1514$, above the predeclared requirement
$N_{\rm eff}\geq1500$.
All Gaussian and Student-$t$ components are normalized over the prior box, so each $q_s$ in
Eq.~\eqref{eq:tsnpe-mis} is its generating density.
The pooled sample contains draws from
$S$ proposal components, each with density $q_s$. Depending on the
composition, this set consists of the TSNPE flow or flows together with the
associated defensive components. These components are combined using the deterministic-mixture
balance heuristic~\cite{Veach:1995mis}. If proposal component $s$ contributes
$n_s$ of the total $N=\sum_s n_s$ draws, then
\begin{equation}
q_{\rm mix}(\bm\theta)
=\sum_{s=1}^{S}\frac{n_s}{N}q_s(\bm\theta),
\qquad
\widetilde w_i=
\frac{\pi(\bm\theta_i)\,
\mathcal L_{\rm full}(\bm\theta_i)}
{q_{\rm mix}(\bm\theta_i)}.
\label{eq:tsnpe-mis}
\end{equation}
After normalization, every quoted TSNPE result reports the resulting
ESS out of the pooled total.
Only the importance-corrected mixture is used for
scientific results; its agreement with UltraNest is established in
Sec.~\ref{sec:results}.

\subsection{Green-function Evidence Network (EN)}
\label{sec:method:en}

We obtain absolute Bayesian evidences with a
Green-function Evidence Network (EN), built in the spirit of Evidence
Networks~\cite{Jeffrey:2023stk}, which amortize Bayesian model comparison over
the data. Here, the amortization is over the nuclear data and the NICER
sources. UltraNest results enter neither the training nor the evaluation of
the EN and are used only for external validation. To the best of our
knowledge, this is the first application of the EN approach to Bayesian
evidence estimation in neutron-star EOS inference. The original Evidence
Network construction trains a classifier with the l-POP--Exponential loss,
whose optimal output is a monotonic function of the Bayes factor. Instead, we
use regression on importance-sampling evidence labels to learn the kernel that maps each
NICER source density to its evidence, as defined below.
All evidence calculations use the unnormalized nuclear-matter
kernel defined in Sec.~\ref{sec:method:like}, following the same convention as
every other inference method. For NICER source $k$, the corresponding
likelihood factor is the average of its mass--radius density $q_k(M,R)$ along
the candidate stellar curve, as described in Sec.~\ref{sec:method:like}. This
dependence is linear in $q_k$, so the evidence can be written as
\begin{equation}
\mathcal Z_k[q_k](\bm x_{\rm nuc})=\int\!\!\int
q_k(M,R)\,G_k(M,R;\bm x_{\rm nuc})\,dM\,dR,
\label{eq:en-green}
\end{equation}
where $G_k$ is obtained by integrating the prior and all remaining likelihood
factors over the EOS parameters, each EOS contributing along its candidate
curve with the same weights as in the NICER average. The remaining factors are
the other two NICER sources, the nuclear-matter kernel at $\bm x_{\rm nuc}$,
GW170817, pQCD, and $M_{\max}$. The function $G_k$ therefore plays the role of
a Green function that maps the source density to its Bayesian evidence.

A single neural network learns $\log G_k(M,R;\bm x_{\rm nuc})$ for
all three sources $k$ on a fixed $(M,R)$ lattice. For nucleonic DDB, the
lattice contains $191\times241$ points spanning $1.0$--$2.9\,\msun$ in mass
and $7$--$19$~km in radius; for hyperonic DDB$\Lambda\Xi^-$, it contains
$161\times205$ points spanning $1.0$--$2.6\,\msun$ and $8.2$--$18.4$~km.
The architecture multiplicatively couples a Fourier-feature coordinate
network to a nuclear-data network.

Training proceeds by regression against evidence labels computed from
banks of exact EOS/TOV/likelihood rows, comprising $848{,}017$ nucleonic rows
and $645{,}084$ hyperonic rows. Each label is the importance-sampling estimate
\begin{equation}
\hat{\mathcal Z}=\frac{\sum_i r_i\,\mathcal L(\bm\theta_i)}{\sum_i r_i},
\label{eq:en-label}
\end{equation}
where $r_i$ is the ratio of the prior density to the sampling intensity of the
bank at row $i$, and the sums run over all bank rows, including EOS without a
valid stellar branch, for which $\mathcal L=0$. The relative Monte Carlo error of each
label is $N_{\rm eff}^{-1/2}$, where $N_{\rm eff}$ is the ESS of the terms
$r_i\mathcal L(\bm\theta_i)$; the denominator, a sum over all ${\sim}10^{6}$
rows, adds a relative fluctuation of only ${\sim}10^{-3}$. The kernel output is fitted with a smooth-$\ell_1$
loss to the log-evidence change relative to the baseline sources, and the
second output with a squared loss to the standardized baseline log evidence. The synthetic NICER sources are single
Gaussians taken from a fixed set of $8192$ scenarios, partitioned into $7168$
for training, $512$ for model selection, and $512$ held out. At every
optimization step, the nuclear data are drawn afresh within $\pm2.5\sigma$ of
the observation. For each EOS model, three independently seeded networks are
trained for $12{,}000$ steps each. Before any query is performed, every network
must reproduce the held-out synthetic sources and the three real baseline
sources within predeclared limits on the mean absolute log-evidence error:
$0.08$--$0.18$ for the families of held-out synthetic sources, and $0.10$ for
the real baseline sources and for the second network output.

The three networks are frozen before any held-out NICER file is
used. They are then queried, without retraining, for the five nuclear settings
and the three held-out sources. For the five nuclear settings, which keep the
three baseline sources, the evidence is given directly by a second output of
the same network, trained on the importance-sampling evidence labels of the baseline sources.
When a held-out source replaces source $k$, its density grid $q_k$, obtained
from the same kernel density estimate used in $\mathcal L_{\rm NICER}$, is
integrated against the frozen kernel $G_k(M,R;\bm x_{\rm nuc})$ through
Eq.~\eqref{eq:en-green}. The reported evidence is the logarithm of the mean
evidence of the three networks. Its quoted uncertainty conservatively combines
the network-to-network scatter, the Monte Carlo error of the
importance-sampling estimate on the bank, and a calibration term derived from
the held-out synthetic sources and the real baseline sources; for a held-out
source, the difference between the network and bank estimates is also
included. Appendix~\ref{app:numerics} tests these uncertainties against
independent, more precise importance-sampling evidences.

\subsection{Validation with unseen NICER sources}
\label{sec:method:validation}
We assess source generalization through a controlled
held-out-source benchmark. In each test, one fiducial NICER likelihood is
replaced by a source absent from A-NET training, while the
other two NICER terms and every non-NICER likelihood factor are held fixed.
Keeping the number of source likelihoods unchanged isolates the response to a
new mass--radius posterior from the separate effect of changing the total
information content of the likelihood. The pairings provide complementary
mass leverage: PSR~J0614$-$3329~\cite{Mauviard:2025dmd} replaces the similarly
massive, canonical-mass PSR~J0437$-$4715; PSR~J1231$-$1411
\cite{Salmi:2024bss} replaces PSR~J0030$+$0451 to extend the test toward lower
masses; and the newly reported PSR~J1614$-$2230 posterior
\cite{Mauviard:2026gzc} replaces the high-mass PSR~J0740$+$6620. These
substitutions are therefore controlled validation interventions, not an
assumption that future observations must replace existing data. In every case,
the checkpoint is frozen before the held-out source is introduced, no
source-specific retraining is performed, and the result is accepted only after
full-likelihood importance correction and the predeclared
importance-sampling gates; it is then compared with an independent UltraNest
calculation.

\section{Results}
\label{sec:results}

\subsection{Nucleonic DDB inference}

\begin{figure*}[t]
  \centering
  \includegraphics[width=0.95\textwidth]{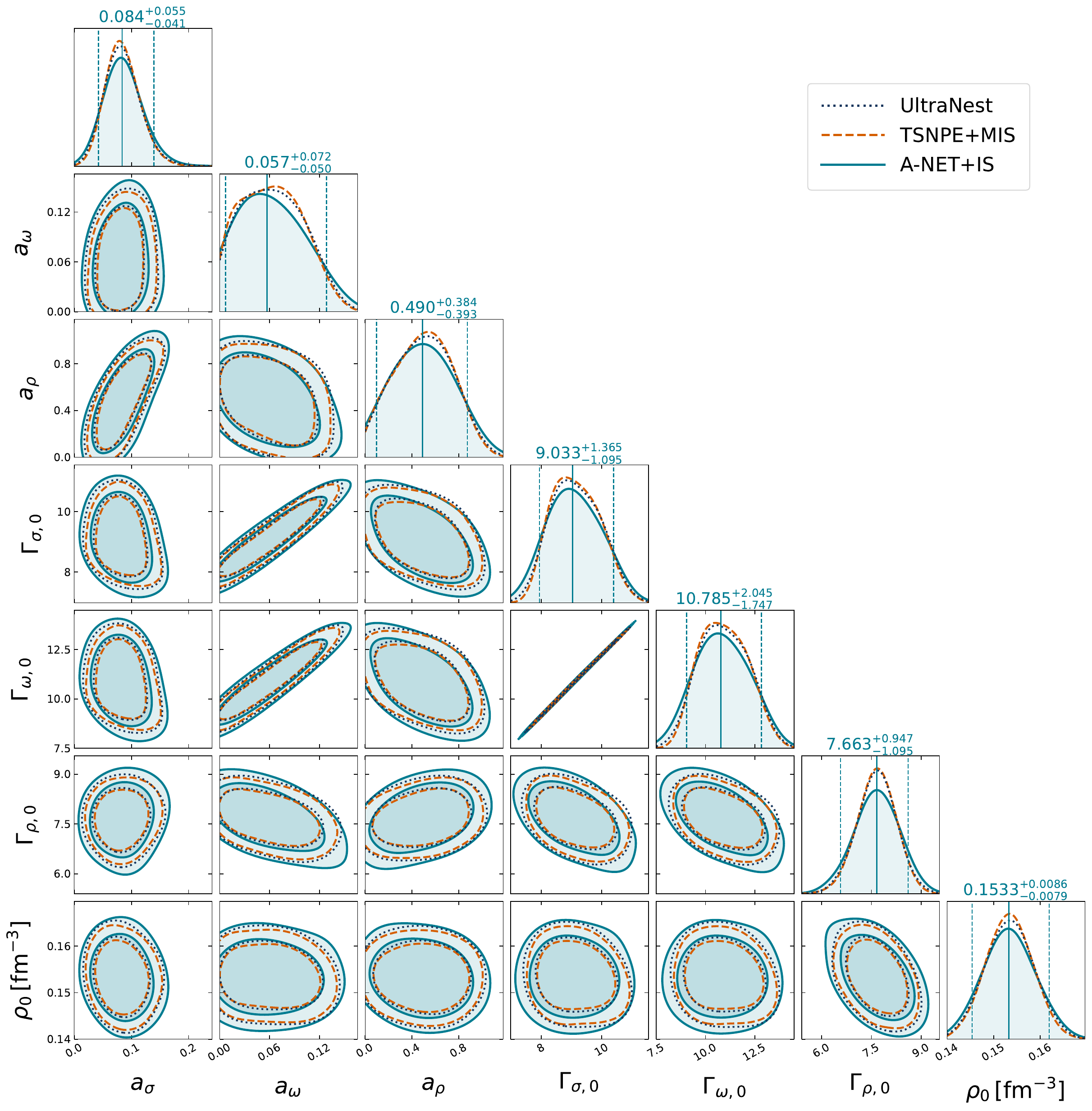}
  \caption{One- and two-dimensional marginal projections of
  the seven-dimensional joint posterior for the nucleonic DDB parameters in
  the fiducial A1 configuration, obtained with UltraNest (navy dotted),
  TSNPE+MIS (orange dashed), and
  A-NET+IS (teal solid
  and shaded). Diagonal panels show the one-dimensional marginal posterior
  densities, while the lower-triangle panels show the 68\% and 90\% credible
  contours of the corresponding two-dimensional marginals. Values above the
  diagonal panels give the A-NET+IS medians and central 90\% credible
  intervals.}
  \label{fig:a1-parameters}
\end{figure*}

Figure~\ref{fig:a1-parameters} compares the one- and
two-dimensional marginal projections of the seven-dimensional posterior for
the fiducial A1 configuration. UltraNest, TSNPE+MIS, and A-NET+IS yield
closely overlapping one-dimensional densities and 68\% and 90\% credible
contours across all DDB parameters. Even the narrow
$\Gamma_{\sigma,0}$--$\Gamma_{\omega,0}$ correlation is recovered with
nearly identical orientation and width. Quantitatively, the largest median
displacement from UltraNest is only $0.7\%$ of the corresponding UltraNest
90\% credible width for TSNPE+MIS and $3.6\%$ for A-NET+IS, while all central
90\% intervals overlap. These results establish two complementary neural
inference routes: TSNPE+MIS provides an accurate and simulation-efficient
sequential analysis targeted to the observation, whereas A-NET+IS achieves
comparable accuracy by reusing a frozen amortized estimator.

\begin{figure*}[t]
  \centering
  \begin{minipage}[t]{0.49\textwidth}
    \centering
    \includegraphics[width=\linewidth]{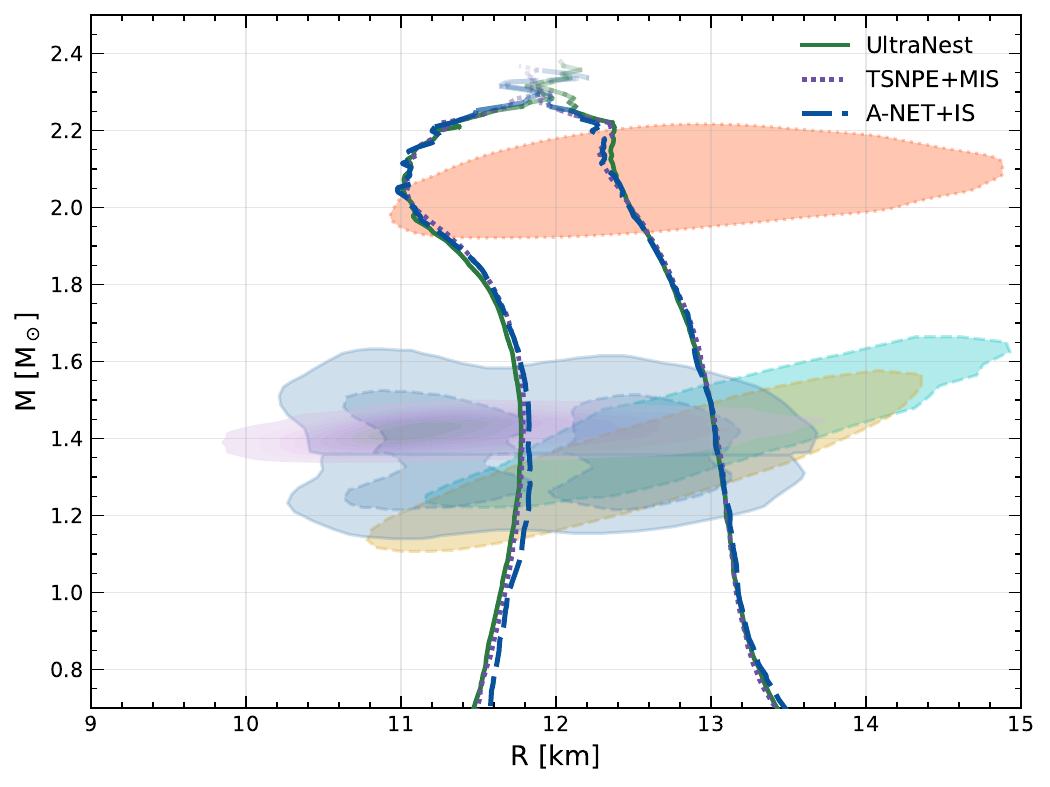}
    \textbf{(a)}
  \end{minipage}\hfill
  \begin{minipage}[t]{0.49\textwidth}
    \centering
    \includegraphics[width=\linewidth]{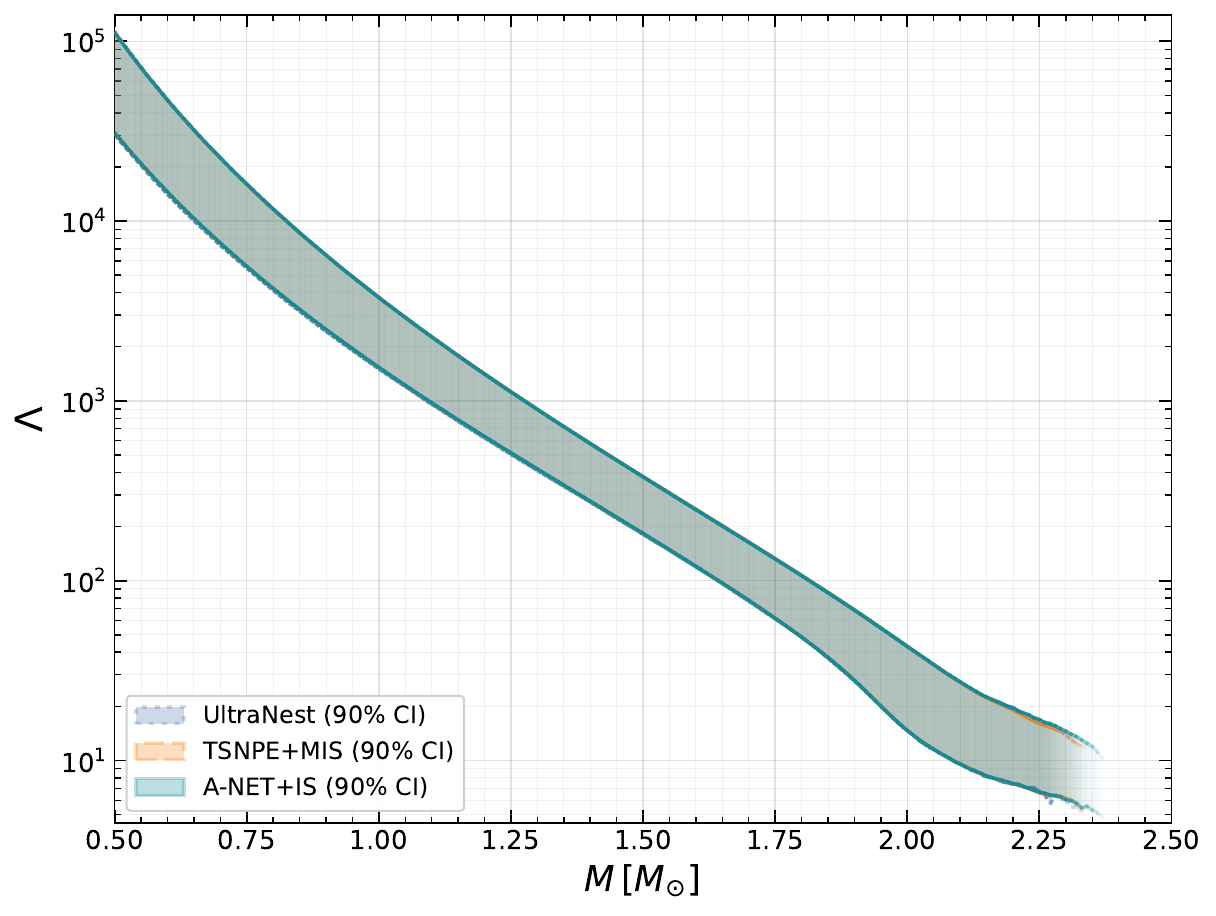}
    \textbf{(b)}
  \end{minipage}
  \caption{90\% credible neutron-star relations for the
  fiducial A1 configuration of the nucleonic DDB model
  \cite{Malik:2022zol}, obtained with UltraNest, TSNPE+MIS, and A-NET+IS.
  \textbf{(a)} Mass--radius bands, with observational constraints from
  PSR~J0030$+$0451
  \cite{Miller:2019cac,Riley:2019yda} (Miller: cyan hatched; Riley: gold),
  PSR~J0740$+$6620 \cite{Dittmann:2024mbo} (orange), PSR~J0437$-$4715
  \cite{Choudhury:2024xbk} (magenta), and
  GW170817 \cite{LIGOScientific:2018cki}
  (steel blue); the likelihood inputs are given in
  Sec.~\ref{sec:method:like}.
  \textbf{(b)} Corresponding mass--tidal-deformability bands for the same
  posterior.}
  \label{fig:a1}
  \label{fig:a1-mass-tidal}
\end{figure*}

\begin{resulttablewide}
  \begin{minipage}{\textwidth}
  \centering
  \caption{Neutron-star observables inferred for the fiducial A1
  configuration of the nucleonic DDB and hyperonic DDB$\Lambda\Xi^-$
  models with UltraNest, TSNPE+MIS, and A-NET+IS. Values are posterior
  medians with 90\% credible intervals. For each composition, the lower
  block gives the mean absolute differences of the lower and upper 90\%
  band edges from UltraNest, averaged over both edges and over evenly spaced
  masses at which both bands are defined (for the mass--tidal relation, with a
  conditional ESS of at least $50$), in km for the mass--radius relation and in
  percent for the mass--tidal relation.}
  \label{tab:a1-summary}
  \footnotesize
  \setlength{\tabcolsep}{2pt}
  \renewcommand{\arraystretch}{1.25}
  \begin{tabular}{@{}lcccc@{}}
    \toprule
    Method & $M_{\max}$ & $R(M_{\max})$ & $R_{1.4}$ & $\Lambda_{1.4}$ \\
    & ($M_\odot$) & (km) & (km) & \\
    \midrule
    \multicolumn{5}{@{}l}{\textit{Nucleonic DDB}} \\
    UltraNest
      & $2.10^{+0.12}_{-0.13}$
      & $11.01^{+0.49}_{-0.51}$
      & $12.41^{+0.62}_{-0.64}$
      & $434^{+144}_{-161}$ \\
    TSNPE+MIS
      & $2.11^{+0.11}_{-0.12}$
      & $11.01^{+0.48}_{-0.49}$
      & $12.43^{+0.61}_{-0.64}$
      & $436^{+145}_{-160}$ \\
    A-NET+IS
      & $2.10^{+0.12}_{-0.12}$
      & $11.00^{+0.48}_{-0.48}$
      & $12.46^{+0.57}_{-0.63}$
      & $436^{+143}_{-159}$ \\
    \addlinespace[2pt]
    \multicolumn{5}{@{}l}{\textit{Hyperonic DDB$\Lambda\Xi^-$}} \\
    UltraNest
      & $2.02^{+0.10}_{-0.10}$
      & $11.47^{+0.52}_{-0.46}$
      & $12.81^{+0.46}_{-0.43}$
      & $574^{+124}_{-113}$ \\
    TSNPE+MIS
      & $2.02^{+0.10}_{-0.10}$
      & $11.46^{+0.49}_{-0.45}$
      & $12.80^{+0.45}_{-0.42}$
      & $570^{+126}_{-107}$ \\
    A-NET+IS
      & $2.02^{+0.09}_{-0.09}$
      & $11.46^{+0.53}_{-0.45}$
      & $12.81^{+0.47}_{-0.43}$
      & $570^{+131}_{-107}$ \\
    \midrule
    \multicolumn{5}{c}{Mean 90\% band-edge difference from UltraNest} \\
    Method
      & \multicolumn{2}{c}{$\langle|\Delta R_{90}|\rangle$ (km)}
      & \multicolumn{2}{c}{$\langle|\delta\Lambda_{90}|\rangle$ (\%)} \\
    \midrule
    \multicolumn{5}{@{}l}{\textit{Nucleonic DDB}} \\
    UltraNest & \multicolumn{2}{c}{---} & \multicolumn{2}{c}{---} \\
    TSNPE+MIS & \multicolumn{2}{c}{$0.033$} & \multicolumn{2}{c}{$1.03$} \\
    A-NET+IS  & \multicolumn{2}{c}{$0.050$} & \multicolumn{2}{c}{$1.13$} \\
    \addlinespace[2pt]
    \multicolumn{5}{@{}l}{\textit{Hyperonic DDB$\Lambda\Xi^-$}} \\
    UltraNest & \multicolumn{2}{c}{---} & \multicolumn{2}{c}{---} \\
    TSNPE+MIS & \multicolumn{2}{c}{$0.024$} & \multicolumn{2}{c}{$1.02$} \\
    A-NET+IS  & \multicolumn{2}{c}{$0.019$} & \multicolumn{2}{c}{$0.86$} \\
    \bottomrule
  \end{tabular}
  \end{minipage}
\end{resulttablewide}

Figure~\ref{fig:a1} compares the 90\% mass--radius
[panel~(a)] and mass--tidal-deformability [panel~(b)] credible bands for the
same fiducial A1 posterior obtained with conventional nested sampling
(UltraNest), TSNPE+MIS, and
A-NET+IS. The three methods closely overlap across their
common mass range in both observables.

For nucleonic DDB,
Table~\ref{tab:a1-summary} quantifies the agreement seen in
Fig.~\ref{fig:a1}: the three posterior medians span only
$0.007\,M_\odot$ in $M_{\max}$, $0.011$~km in $R(M_{\max})$, and
$0.042$~km in $R_{1.4}$, all far smaller than their corresponding 90\%
credible widths. The mean absolute mass--radius band-edge separations
from UltraNest are $0.033$~km for TSNPE+MIS and $0.050$~km for A-NET+IS
(Table~\ref{tab:a1-summary}). Panel~(b) shows equally close agreement for the
tidal deformability. The UltraNest band in this panel combines two independent
UltraNest runs with equal weight. UltraNest, TSNPE+MIS, and A-NET+IS give
$\Lambda_{1.4}=434$, $436$, and $436$, respectively, with strongly overlapping
central 90\% intervals. The mean absolute tidal-band-edge differences from
UltraNest are only $1.03\%$ for TSNPE+MIS and $1.13\%$ for A-NET+IS. In particular, the agreement of A-NET+IS with
conventional nested sampling shows that, for A1, a frozen model amortized over
source configurations preserves posterior accuracy, while TSNPE+MIS supplies
an independent sequential-neural cross-check.
Together, these comparisons show that the agreement
extends from the EOS-parameter posterior to both derived stellar relations.

\begin{resultfigurewide}
  \centering
  \includegraphics[width=0.84\textwidth]{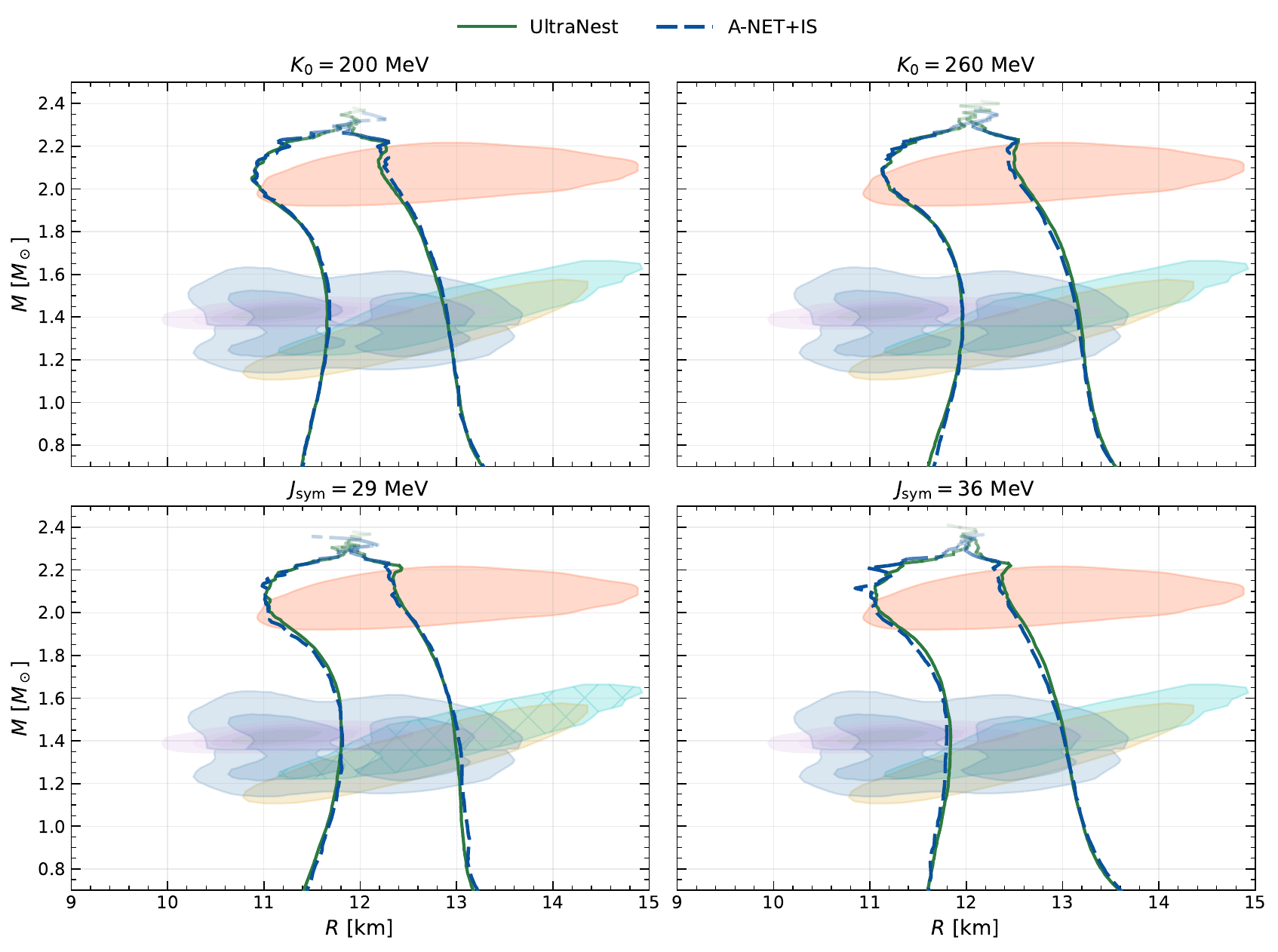}
  \caption{90\% credible mass--radius bands for four shifted
  nuclear-data configurations obtained with UltraNest (solid green) and
  A-NET+IS (dashed blue). Observational constraints as
  in Fig.~\ref{fig:a1}.}
  \label{fig:nuclear-amortization}
  \par\vspace{4pt}
  \begin{minipage}{0.98\textwidth}
    \centering
    \makeatletter\def\@captype{table}\makeatother
    \caption{Posterior medians and central 90\% credible intervals for the
    neutron-star observables in the four shifted nuclear-data configurations of
    Fig.~\ref{fig:nuclear-amortization}. $M_{\max}$ is in $M_\odot$;
    $R(M_{\max})$ and $R_{1.4}$ are in km.}
    \label{tab:nuclear-amortization-summary}
    \footnotesize
    \setlength{\tabcolsep}{10pt}
    \renewcommand{\arraystretch}{1.18}
    \begin{tabular}{@{}llccc@{}}
      \toprule
      Configuration & Method & $M_{\max}$ & $R(M_{\max})$ & $R_{1.4}$ \\
      \midrule
      $K_0=200$ & UltraNest
        & $2.10^{+0.12}_{-0.12}$
        & $10.91^{+0.47}_{-0.47}$
        & $12.27^{+0.62}_{-0.62}$ \\
      & A-NET+IS
        & $2.11^{+0.12}_{-0.13}$
        & $10.86^{+0.52}_{-0.42}$
        & $12.26^{+0.64}_{-0.58}$ \\
      \addlinespace[2pt]
      $K_0=260$ & UltraNest
        & $2.11^{+0.12}_{-0.13}$
        & $11.14^{+0.48}_{-0.53}$
        & $12.57^{+0.61}_{-0.61}$ \\
      & A-NET+IS
        & $2.10^{+0.12}_{-0.12}$
        & $11.12^{+0.46}_{-0.49}$
        & $12.57^{+0.59}_{-0.60}$ \\
      \addlinespace[2pt]
      $J_{\mathrm{sym}}=29$ & UltraNest
        & $2.10^{+0.12}_{-0.12}$
        & $11.00^{+0.48}_{-0.50}$
        & $12.44^{+0.55}_{-0.64}$ \\
      & A-NET+IS
        & $2.10^{+0.12}_{-0.12}$
        & $10.97^{+0.51}_{-0.49}$
        & $12.43^{+0.60}_{-0.61}$ \\
      \addlinespace[2pt]
      $J_{\mathrm{sym}}=36$ & UltraNest
        & $2.11^{+0.12}_{-0.12}$
        & $11.05^{+0.47}_{-0.49}$
        & $12.44^{+0.60}_{-0.60}$ \\
      & A-NET+IS
        & $2.11^{+0.13}_{-0.12}$
        & $11.02^{+0.46}_{-0.50}$
        & $12.44^{+0.58}_{-0.64}$ \\
      \bottomrule
    \end{tabular}
  \end{minipage}
\end{resultfigurewide}

Figure~\ref{fig:nuclear-amortization} tests nuclear-data
amortization by changing one nuclear-matter measurement at a time while
keeping all astrophysical likelihood terms fixed. Without retraining,
A-NET+IS reproduces the UltraNest 90\% mass--radius bands in all four
configurations. The mean absolute band-edge separations are only $0.022$,
$0.029$, $0.034$, and $0.036$~km for $K_0=200$, $K_0=260$,
$J_{\mathrm{sym}}=29$, and $J_{\mathrm{sym}}=36$~MeV, respectively.
Table~\ref{tab:nuclear-amortization-summary} confirms the same agreement for
the scalar observables: across all configurations, the largest differences
between the posterior medians are $0.006\,M_\odot$ in $M_{\max}$, $0.046$~km
in $R(M_{\max})$, and $0.013$~km in $R_{1.4}$, with overlapping central 90\%
intervals in every case. Increasing $K_0$ from $200$ to $260$~MeV shifts
$R_{1.4}$ upward by approximately $0.30$~km in both methods, demonstrating
that the frozen network follows the posterior response to the nuclear input
rather than merely reproducing the fiducial A1 result. Thus, A-NET+IS extends
amortization to nuclear measurements while retaining agreement with
conventional Bayesian inference.

\begin{resultfigurewide}
  \centering
  \includegraphics[width=0.49\linewidth]{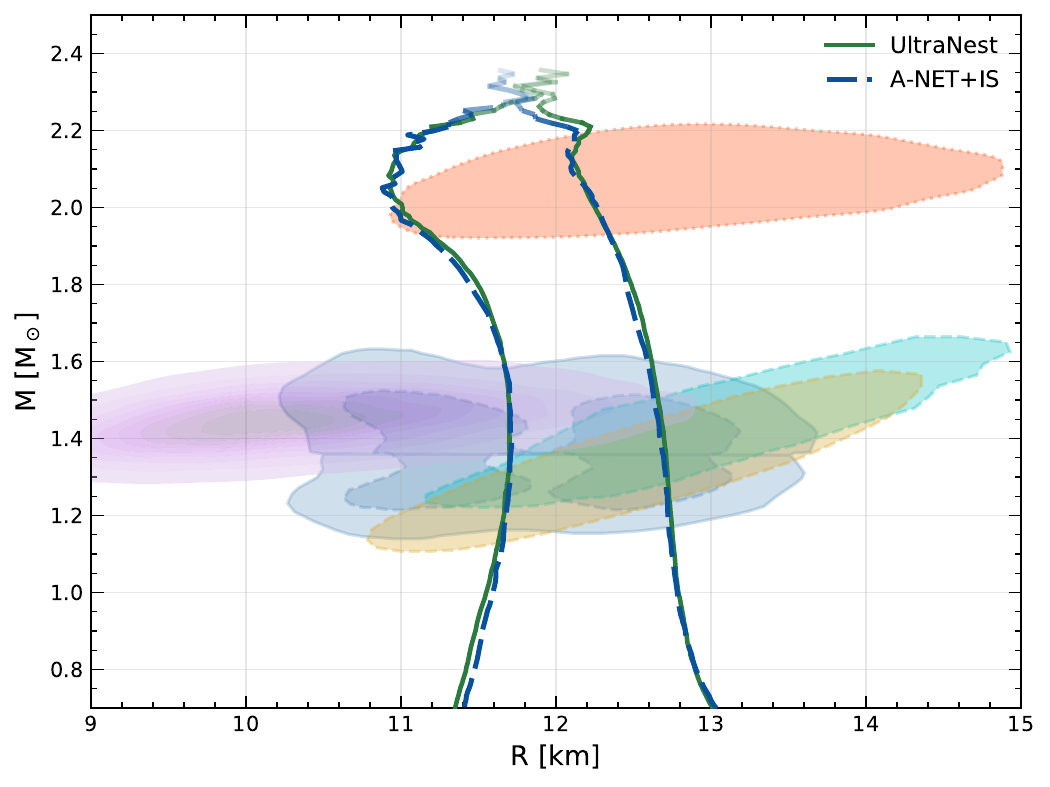}\hfill
  \includegraphics[width=0.49\linewidth]{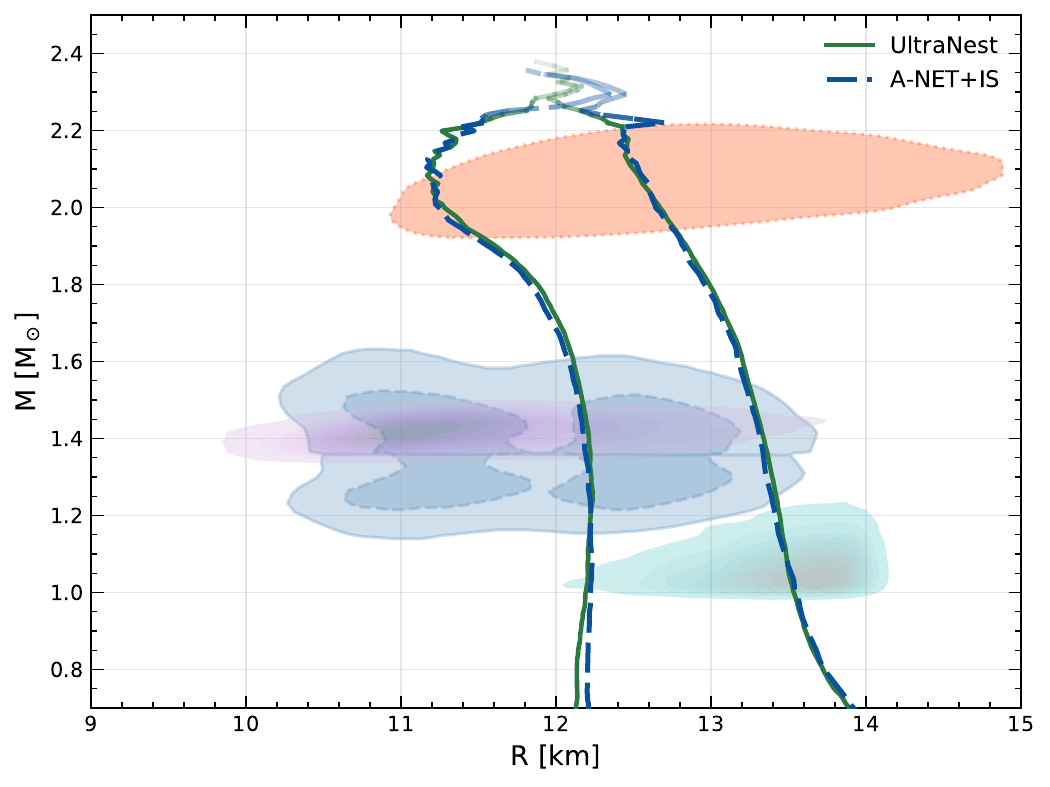}
  \caption{90\% credible mass--radius bands for two substituted-source
  configurations, obtained with UltraNest (solid green) and
  A-NET+IS (dashed blue).
  Left: PSR~J0614$-$3329 \cite{Mauviard:2025dmd} replaces
  PSR~J0437$-$4715 and is violet; retained PSR~J0030$+$0451 is cyan
  hatched/gold. Right: PSR~J1231$-$1411 \cite{Salmi:2024bss} replaces
  PSR~J0030$+$0451 and is teal; retained PSR~J0437$-$4715 is magenta.
  In both panels, retained PSR~J0740$+$6620 is orange and GW170817 is
  steel blue; baseline-source references are given in Fig.~\ref{fig:a1}.}
  \label{fig:substituted}
  \par\vspace{4pt}
  \begin{minipage}{0.98\textwidth}
    \centering
    \makeatletter\def\@captype{table}\makeatother
    \caption{Posterior medians and central 90\% credible intervals for the
    neutron-star observables in the three substituted-source configurations of
    Figs.~\ref{fig:substituted} and~\ref{fig:j1614-substituted}
    (nucleonic) and Figs.~\ref{fig:hyp-substituted}
    and~\ref{fig:hyp-j1614-substituted} (hyperonic). $M_{\max}$ is in $M_\odot$;
    $R(M_{\max})$ and $R_{1.4}$ are in km.}
    \label{tab:substituted-summary}
    \scriptsize
    \setlength{\tabcolsep}{4.5pt}
    \renewcommand{\arraystretch}{1.18}
    \begin{tabular}{@{}llcccccc@{}}
      \toprule
      & & \multicolumn{3}{c}{\textit{Nucleonic DDB}}
        & \multicolumn{3}{c}{\textit{Hyperonic DDB$\Lambda\Xi^-$}} \\
      \cmidrule(lr){3-5}\cmidrule(l){6-8}
      Configuration & Method & $M_{\max}$ & $R(M_{\max})$ & $R_{1.4}$
        & $M_{\max}$ & $R(M_{\max})$ & $R_{1.4}$ \\
      \midrule
      PSR~J0614$-$3329 & UltraNest
        & $2.10^{+0.11}_{-0.12}$ & $10.91^{+0.41}_{-0.43}$ & $12.28^{+0.42}_{-0.58}$
        & $1.99^{+0.08}_{-0.08}$ & $11.22^{+0.25}_{-0.37}$ & $12.55^{+0.16}_{-0.33}$ \\
       & A-NET+IS
        & $2.10^{+0.11}_{-0.12}$ & $10.88^{+0.46}_{-0.44}$ & $12.28^{+0.39}_{-0.57}$
        & $1.99^{+0.07}_{-0.08}$ & $11.22^{+0.27}_{-0.35}$ & $12.55^{+0.17}_{-0.29}$ \\
      \addlinespace[2pt]
      PSR~J1231$-$1411 & UltraNest
        & $2.09^{+0.12}_{-0.12}$ & $11.15^{+0.47}_{-0.46}$ & $12.76^{+0.58}_{-0.55}$
        & $2.02^{+0.10}_{-0.10}$ & $11.66^{+0.51}_{-0.48}$ & $13.02^{+0.42}_{-0.45}$ \\
       & A-NET+IS
        & $2.09^{+0.13}_{-0.12}$ & $11.14^{+0.45}_{-0.46}$ & $12.78^{+0.54}_{-0.59}$
        & $2.03^{+0.09}_{-0.10}$ & $11.65^{+0.53}_{-0.46}$ & $13.02^{+0.43}_{-0.45}$ \\
      \addlinespace[2pt]
      PSR~J1614$-$2230 & UltraNest
        & $2.02^{+0.14}_{-0.09}$ & $10.72^{+0.56}_{-0.48}$ & $12.19^{+0.68}_{-0.71}$
        & $1.97^{+0.10}_{-0.05}$ & $11.30^{+0.56}_{-0.44}$ & $12.68^{+0.50}_{-0.43}$ \\
       & A-NET+IS
        & $2.02^{+0.14}_{-0.08}$ & $10.70^{+0.57}_{-0.44}$ & $12.18^{+0.69}_{-0.67}$
        & $1.97^{+0.10}_{-0.05}$ & $11.30^{+0.56}_{-0.44}$ & $12.67^{+0.49}_{-0.43}$ \\
      \bottomrule
    \end{tabular}
  \end{minipage}
\end{resultfigurewide}

A-NET conditions on exactly three NICER sources, so each
unseen source enters by replacing one fiducial source while all other
likelihood terms are kept fixed. This keeps the number of NICER likelihood
terms unchanged and isolates the response to the new mass--radius posterior
(Sec.~\ref{sec:method:validation}); an additional source could instead be
added through the importance weights, subject to the ESS gate. The left panel
of Fig.~\ref{fig:substituted} tests A-NET on PSR~J0614$-$3329,
which was not included in training and replaces
PSR~J0437$-$4715. Its preference for
smaller radii moves the posterior to $R_{1.4}=12.28$~km with both UltraNest and
A-NET+IS. The two 90\% mass--radius bands remain closely aligned, with a mean
absolute band-edge separation of $0.046$~km over their common mass range.
Table~\ref{tab:substituted-summary} gives the same conclusion for the derived
observables: the posterior medians differ by only $0.004\,M_\odot$ in
$M_{\max}$, $0.028$~km in $R(M_{\max})$, and $0.007$~km in $R_{1.4}$, and all
central 90\% intervals overlap. The frozen network therefore follows the
radius information supplied by the unseen source without retraining.

The right panel provides a complementary test with
PSR~J1231$-$1411 replacing PSR~J0030$+$0451. The new source probes the
low-mass part of the relation and favors larger radii, shifting the inferred
$R_{1.4}$ upward to $12.76$~km with UltraNest and $12.78$~km with A-NET+IS.
Despite this response being opposite to that induced by PSR~J0614$-$3329, the
two methods again recover nearly identical 90\% bands: their mean absolute
band-edge separation is $0.045$~km, while their median differences are only
$0.001\,M_\odot$, $0.007$~km, and $0.014$~km in $M_{\max}$,
$R(M_{\max})$, and $R_{1.4}$, respectively. Taken together, the two panels are
the central demonstration of astrophysical amortization in this work. The same
frozen A-NET assimilates two source likelihoods absent from training, resolves
their distinct data-driven posterior shifts, and, after full-likelihood
IS, reproduces independent UltraNest inferences without
retraining.

\begin{resultfigurewide}
  \begin{minipage}[c]{0.48\textwidth}
    \centering
    \includegraphics[width=\linewidth]{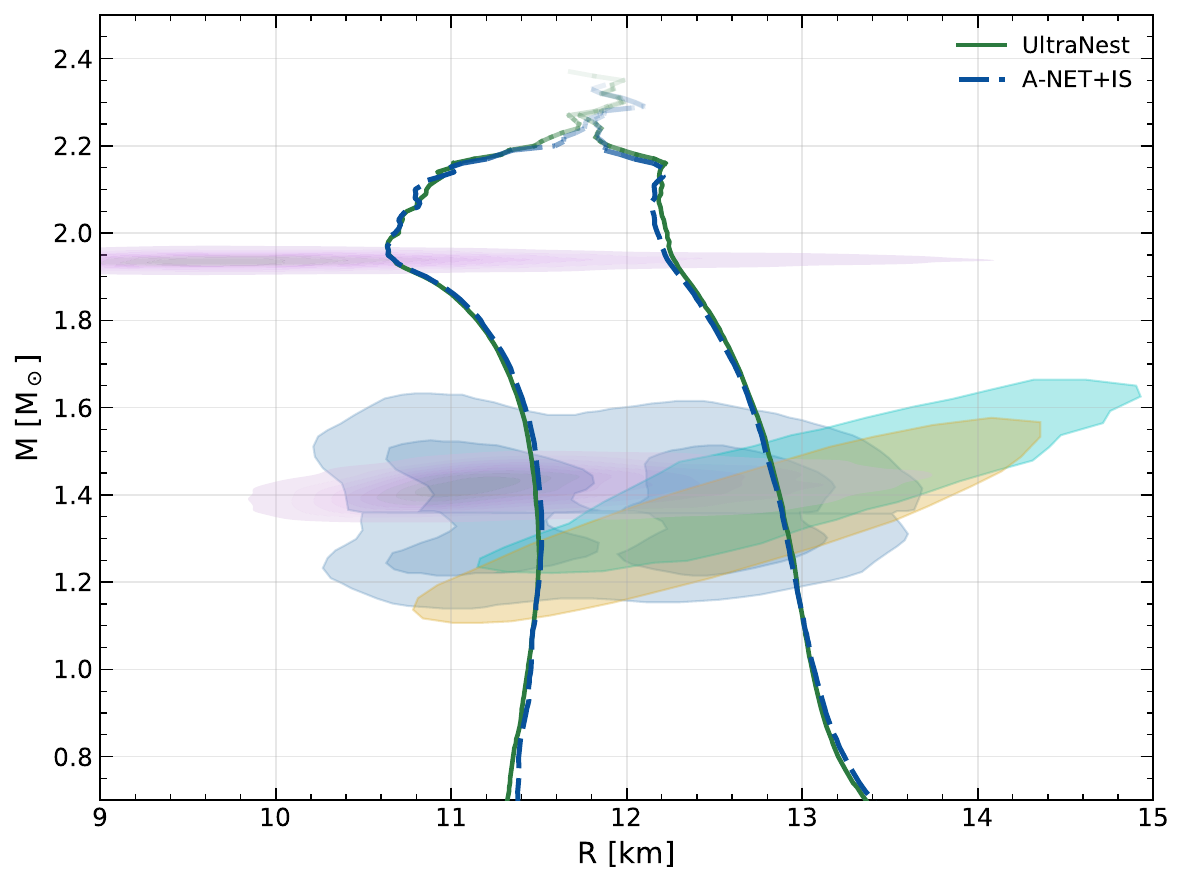}
  \end{minipage}\hfill
  \begin{minipage}[c]{0.48\textwidth}
    \caption{90\% credible mass--radius bands for the
    PSR~J1614$-$2230 substituted-source configuration, obtained with UltraNest
    (solid green) and A-NET+IS (dashed blue).
    PSR~J1614$-$2230 replaces PSR~J0740$+$6620. Observational
    constraints: PSR~J1614$-$2230
    \cite{Mauviard:2026gzc} (violet), retained PSR~J0030$+$0451 (cyan
    hatched/gold), retained PSR~J0437$-$4715 (lighter magenta), and GW170817
    (steel blue); baseline-source references are given in Fig.~\ref{fig:a1}.}
    \label{fig:j1614-substituted}
  \end{minipage}
\end{resultfigurewide}

Figure~\ref{fig:j1614-substituted} extends the unseen-source
test to the newly reported high-mass, compact PSR~J1614$-$2230 posterior. Its
preference for smaller radii shifts the inference to
$R_{1.4}=12.19$~km with UltraNest and $12.18$~km with A-NET+IS. The two
90\% mass--radius bands overlap throughout their common reliable mass range
(masses reached by at least $5\%$ of the posterior weight with an ESS of at
least $100$), with a mean absolute band-edge separation of only $0.027$~km.
Table~\ref{tab:substituted-summary} shows equally close agreement in the
derived observables: the posterior medians differ by only
$0.004\,M_\odot$ in $M_{\max}$, $0.019$~km in $R(M_{\max})$, and
$0.010$~km in $R_{1.4}$, with overlapping central 90\% intervals. Thus the
same frozen A-NET, without retraining, also follows an unseen compact-star
likelihood in the high-mass NICER slot and reproduces the independent
UltraNest inference after IS.

\subsection{\texorpdfstring{Hyperonic
DDB$\Lambda\Xi^-$ inference}{Hyperonic DDB Lambda Xi-minus inference}}

To test composition dependence, we repeat
the complete analysis for nine-dimensional hyperonic DDB$\Lambda\Xi^-$,
using SU(6)-fixed vector couplings (including the
hidden-strangeness $\phi$ meson) and uniform, hypernuclear-calibrated priors
$x_{\sigma\Lambda}\in[0.609,0.622]$ and
$x_{\sigma\Xi}\in[0.309,0.322]$
\cite{Malik:2022jqc,Fortin:2017cvt,Fortin:2020qin}.
The crust, likelihood, and data remain
otherwise unchanged.

\begin{resultfigurewide}
  \centering
  \begin{minipage}[t]{0.49\textwidth}
    \centering
    \includegraphics[width=\linewidth]{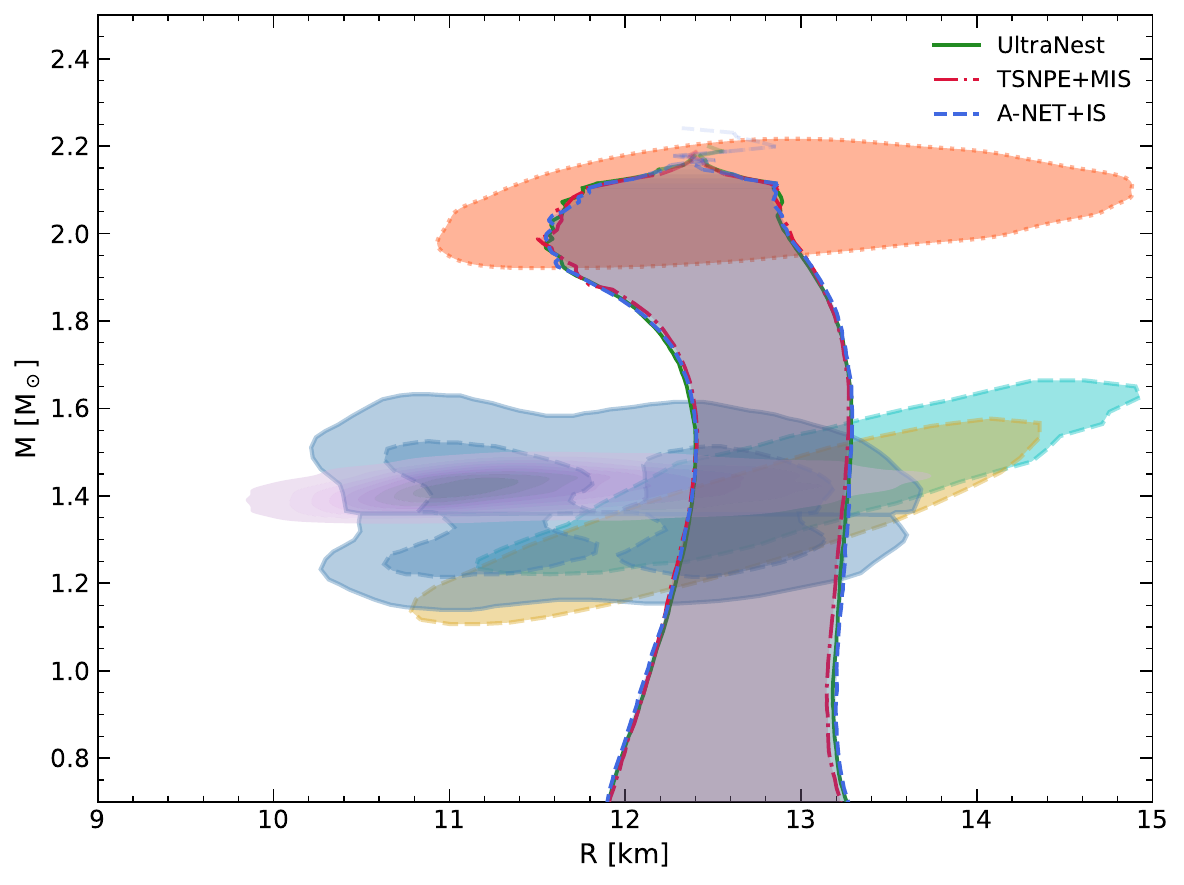}
    \textbf{(a)}
  \end{minipage}\hfill
  \begin{minipage}[t]{0.49\textwidth}
    \centering
    \includegraphics[width=\linewidth]{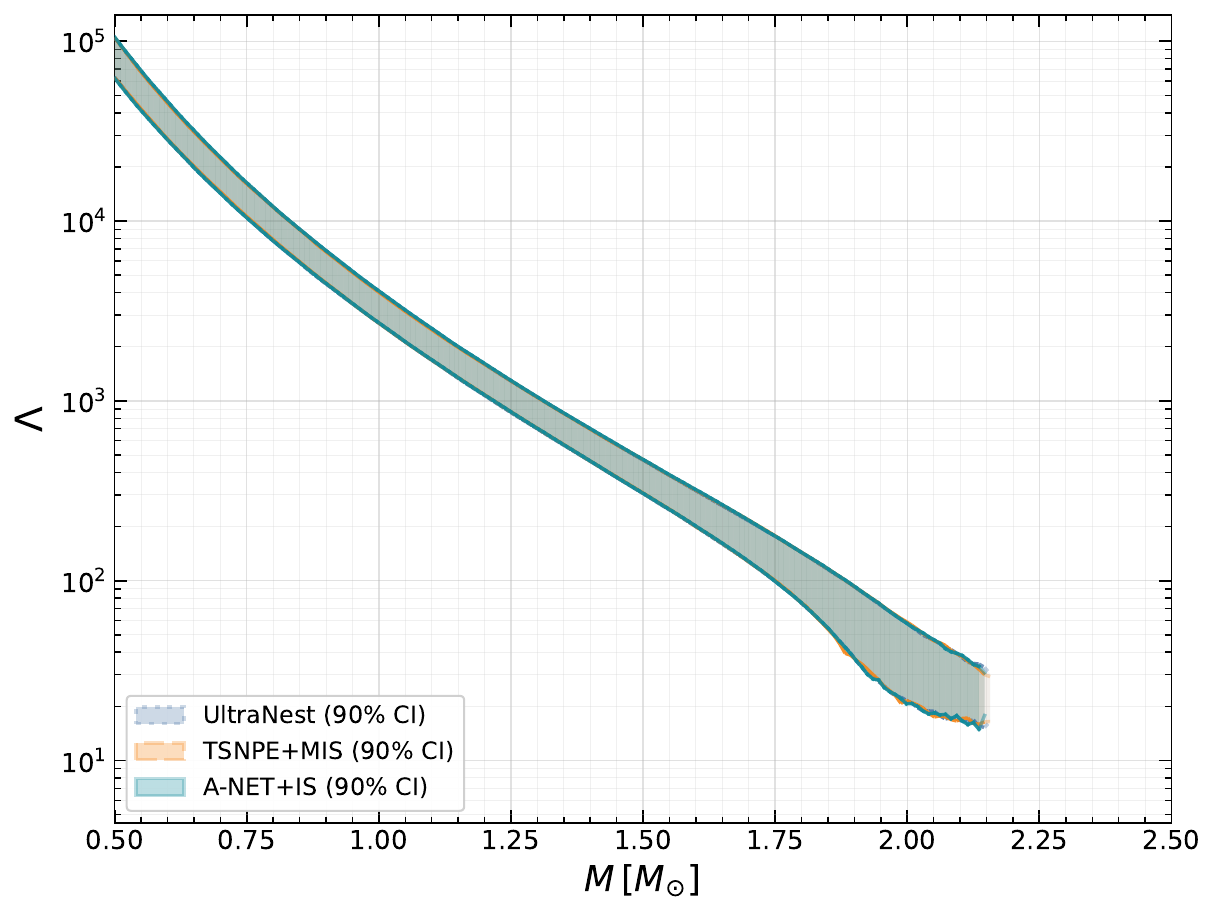}
    \textbf{(b)}
  \end{minipage}
  \caption{90\% credible neutron-star relations for the
  fiducial A1 configuration of the hyperonic DDB$\Lambda\Xi^-$ model,
  obtained with UltraNest, TSNPE+MIS, and A-NET+IS. \textbf{(a)}
  Mass--radius bands; observational constraints as in
  Fig.~\ref{fig:a1}. \textbf{(b)} Corresponding
  mass--tidal-deformability bands for the same posterior.}
  \label{fig:hyp-a1}
  \label{fig:hyp-a1-mass-tidal}
\end{resultfigurewide}

\begin{resultfigurewide}
  \begin{minipage}{\textwidth}
    \centering
    \includegraphics[width=0.49\linewidth]{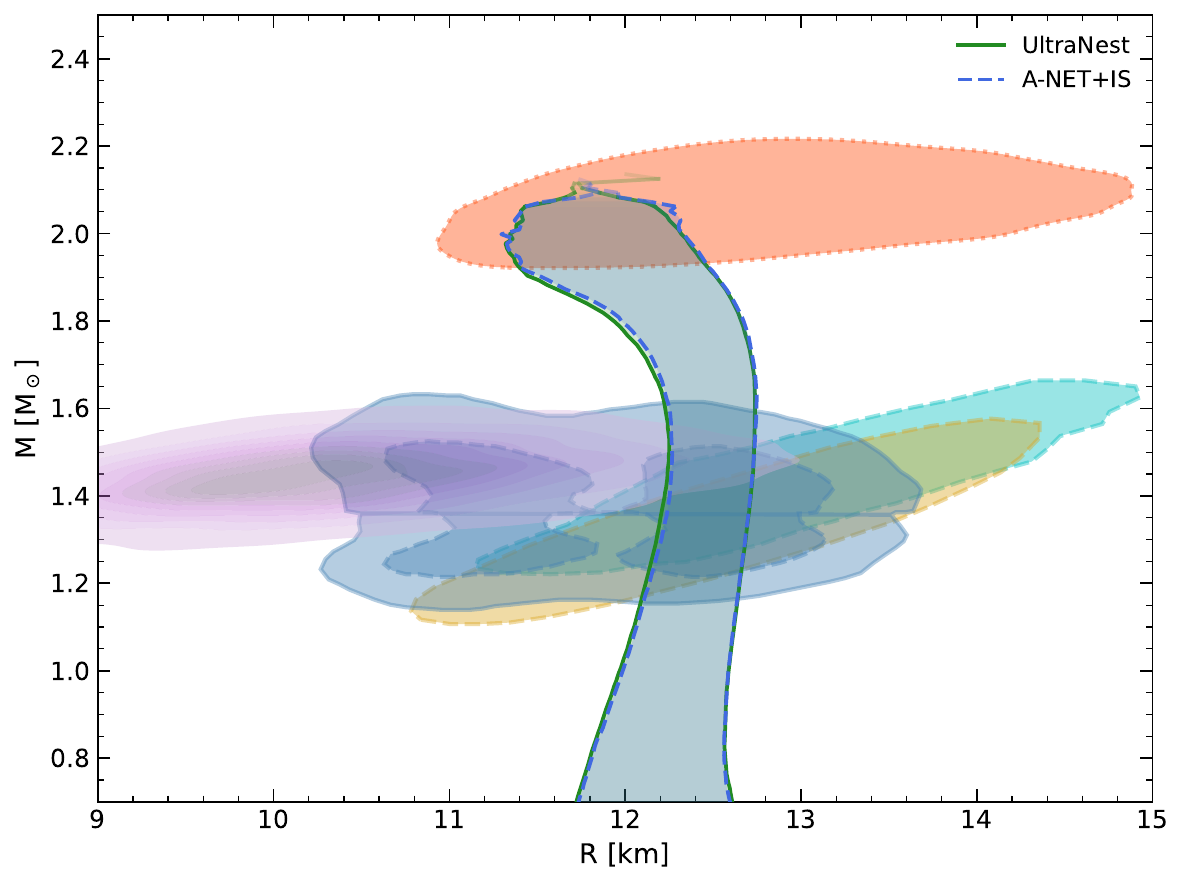}\hfill
    \includegraphics[width=0.49\linewidth]{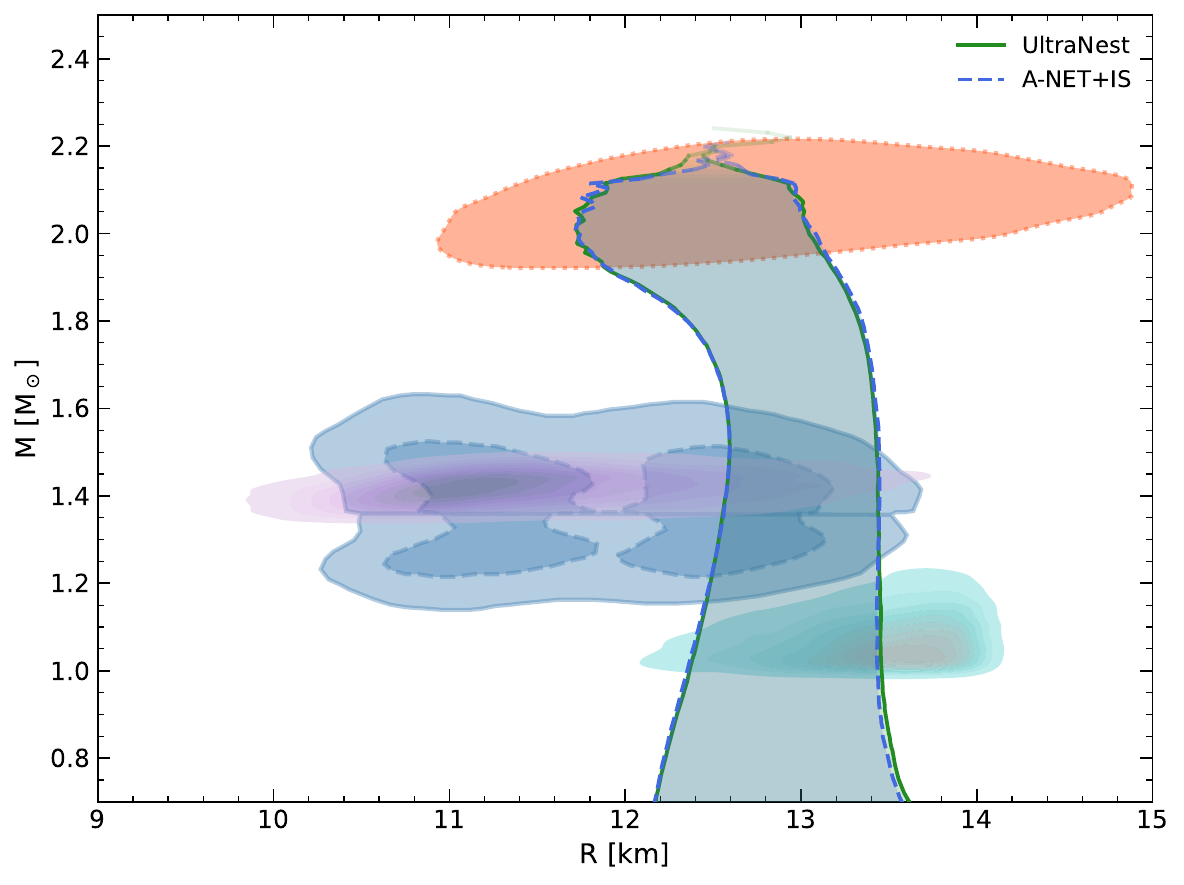}
    \caption{90\% credible mass--radius bands for the hyperonic
    DDB$\Lambda\Xi^-$ model under two NICER source substitutions, inferred
    with UltraNest (solid green) and A-NET+IS (dashed blue).
    Left: PSR~J0614$-$3329 \cite{Mauviard:2025dmd} replaces
    PSR~J0437$-$4715 and is magenta; retained
    PSR~J0030$+$0451 is cyan hatched/gold. Right: PSR~J1231$-$1411
    \cite{Salmi:2024bss} replaces PSR~J0030$+$0451 and is teal; retained
    PSR~J0437$-$4715 is magenta. In both panels, retained
    PSR~J0740$+$6620 is orange and GW170817 is steel blue;
    baseline-source references are given in Fig.~\ref{fig:a1}.}
    \label{fig:hyp-substituted}
  \end{minipage}
\end{resultfigurewide}

Figure~\ref{fig:hyp-a1} compares the 90\%
mass--radius and mass--tidal-deformability bands for the fiducial hyperonic A1
posterior. Both neural methods closely overlap UltraNest in each panel, as
quantified by the hyperonic block of Table~\ref{tab:a1-summary}. In panel~(b),
UltraNest, TSNPE+MIS, and A-NET+IS give $\Lambda_{1.4}=574$, $570$, and $570$,
respectively. Their 90\% tidal bands overlap throughout the common reliable
mass range, with mean tidal-band-edge differences from UltraNest of $1.02\%$
for TSNPE+MIS and $0.86\%$ for A-NET+IS. The corresponding mean
mass--radius-band-edge differences are $0.024$~km and $0.019$~km,
respectively. The close agreement among all three methods therefore persists
after enlarging both the EOS composition and the parameter space.

Relative to nucleonic A1, the hyperonic posterior has a
larger median $R_{1.4}$ and $\Lambda_{1.4}$ but a lower median $M_{\max}$. The
larger $R_{1.4}$ and $\Lambda_{1.4}$ agree in direction with
Ref.~\cite{Malik:2022jqc}: when hyperons are present, the
${\sim}2\,M_\odot$ constraint requires a stiffer isoscalar (nucleonic) sector,
which produces larger radii and tidal deformabilities. The lower median
$M_{\max}$ reflects the softening after hyperons appear at high density.

\begin{resultfigurewide}
  \begin{minipage}[c]{0.48\textwidth}
    \centering
    \includegraphics[width=\linewidth]{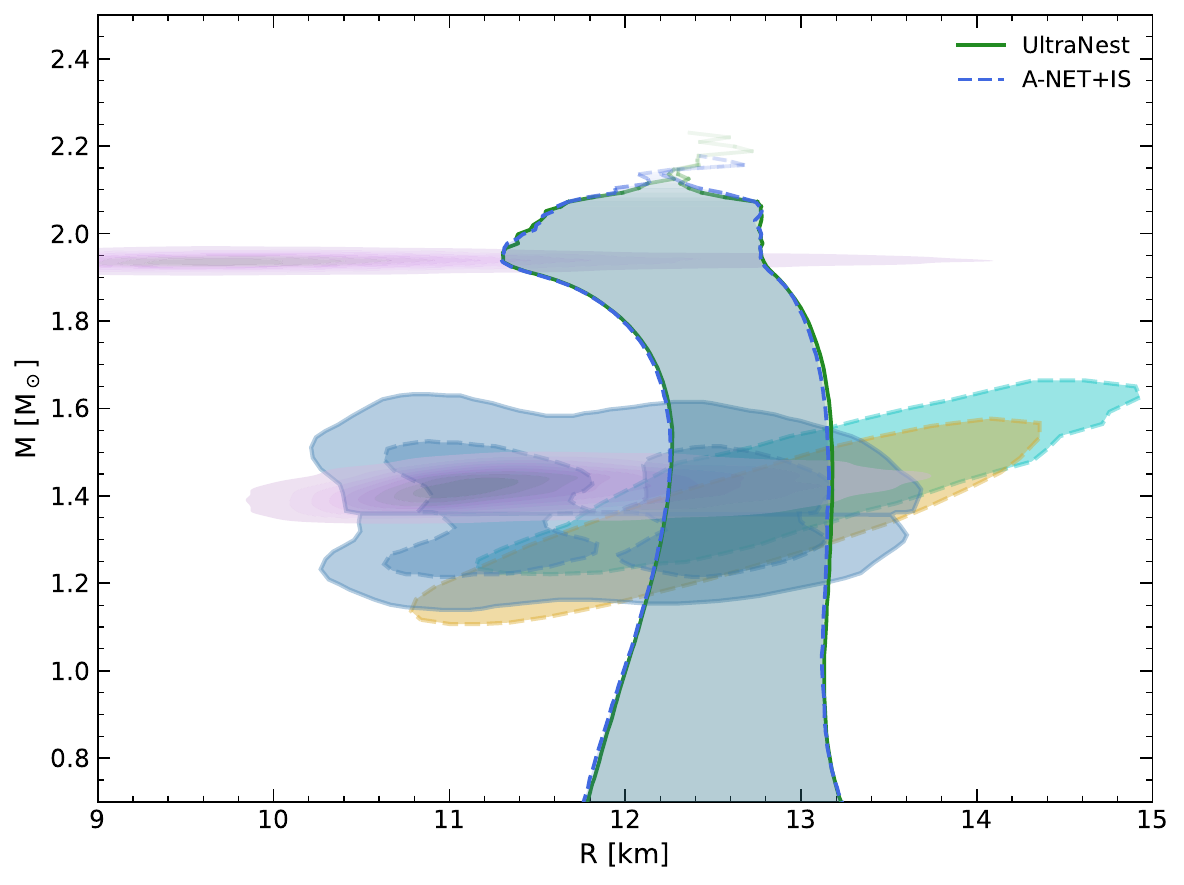}
  \end{minipage}\hfill
  \begin{minipage}[c]{0.48\textwidth}
    \caption{90\% credible mass--radius bands for the
    PSR~J1614$-$2230 substituted-source configuration of the hyperonic
    DDB$\Lambda\Xi^-$ model, obtained with UltraNest (solid green) and
    A-NET+IS (dashed blue). PSR~J1614$-$2230
    replaces PSR~J0740$+$6620. Observational constraints:
    PSR~J1614$-$2230
    \cite{Mauviard:2026gzc} (violet), retained PSR~J0030$+$0451 (cyan
    hatched/gold), retained PSR~J0437$-$4715 (lighter magenta), and GW170817
    (steel blue); baseline-source references are given in
    Fig.~\ref{fig:a1}.}
    \label{fig:hyp-j1614-substituted}
  \end{minipage}
\end{resultfigurewide}

Figure~\ref{fig:hyp-substituted} repeats the two
unseen-source tests for the hyperonic model; both sources were absent from
A-NET training. For PSR~J0614$-$3329, UltraNest gives
$R_{1.4}=12.550$~km and $M_{\max}=1.990\,M_\odot$, while A-NET+IS gives
$12.547$~km and $1.991\,M_\odot$. The mean 90\% band-edge separation is
$0.016$~km, with a maximum of $0.118$~km over the common reliable mass
range. For PSR~J1231$-$1411, UltraNest and A-NET+IS both give
$R_{1.4}=13.019$~km, with $M_{\max}=2.024$ and $2.029\,M_\odot$,
respectively. The mean band-edge separation is $0.016$~km, with a maximum of
$0.089$~km. For both sources, the 90\% intervals overlap at every mass point
in the common reliable range. Table~\ref{tab:substituted-summary} gives the
same agreement for the derived observables: for both sources, all central
90\% intervals of $M_{\max}$, $R(M_{\max})$, and $R_{1.4}$ overlap, and the
$R(M_{\max})$ medians differ by only $0.007$ and $0.009$~km, respectively.
The frozen hyperonic A-NET follows both held-out source likelihoods without
target-specific retraining and reproduces their distinct UltraNest posterior
shifts after IS.

Figure~\ref{fig:hyp-j1614-substituted} extends the
hyperonic held-out-source test to PSR~J1614$-$2230. The UltraNest and
A-NET+IS 90\% bands overlap throughout their common reliable range, with a
mean band-edge separation of $0.013$~km. Table~\ref{tab:substituted-summary}
shows the same agreement in the derived observables: the posterior medians of
$M_{\max}$ and $R(M_{\max})$ agree within their Monte Carlo uncertainties of
about $0.0007\,M_\odot$ and $0.006$~km, those of $R_{1.4}$ differ by
$0.009$~km, and the central 90\% intervals overlap. After IS, A-NET reproduces the UltraNest result without
target-specific retraining.

\subsection{Green-function Evidence Network (EN)}
\label{sec:results:evidence-network}

\begin{resulttablewide}
  \begin{minipage}{\textwidth}
  \caption{Bayesian-evidence comparison between the Green-function Evidence
  Network (EN) and UltraNest for eight configurations of each of the
  nucleonic DDB and hyperonic DDB$\Lambda\Xi^-$ models. The two central
  columns report the log evidence $\log\mathcal{Z}$ with its quoted $1\sigma$
  uncertainty; the final column gives the EN--UltraNest separation in units
  of their combined uncertainty, so values below 1 mean agreement within
  $1\sigma$. $K_0$ and $J_{\mathrm{sym}}$ are in MeV.}
  \label{tab:en-ultranest}
  \label{tab:en-ultranest-hyp}
  \centering
  \footnotesize
  \setlength{\tabcolsep}{5pt}
  \renewcommand{\arraystretch}{0.90}
  \begin{tabular*}{\textwidth}{@{\extracolsep{\fill}}lccc@{}}
    \toprule
    Configuration & EN & UltraNest & Separation ($\sigma$) \\
    \midrule
    \multicolumn{4}{@{}l}{\textit{Nucleonic DDB}} \\
    A1 base                         & $-25.613 \pm 0.050$ & $-25.467 \pm 0.242$ & $0.59$ \\
    $K_0=200$                       & $-25.582 \pm 0.053$ & $-25.631 \pm 0.215$ & $0.22$ \\
    $K_0=260$                       & $-25.851 \pm 0.048$ & $-25.865 \pm 0.160$ & $0.08$ \\
    $J_{\mathrm{sym}}=29$          & $-25.542 \pm 0.127$ & $-25.596 \pm 0.106$ & $0.33$ \\
    $J_{\mathrm{sym}}=36$          & $-25.721 \pm 0.114$ & $-25.766 \pm 0.102$ & $0.29$ \\
    PSR~J0614$-$3329                & $-27.170 \pm 0.166$ & $-26.972 \pm 0.160$ & $0.85$ \\
    PSR~J1231$-$1411                & $-26.286 \pm 0.234$ & $-26.448 \pm 0.213$ & $0.51$ \\
    PSR~J1614$-$2230                & $-25.585 \pm 0.179$ & $-25.565 \pm 0.286$ & $0.06$ \\
    \addlinespace[1pt]
    \multicolumn{4}{@{}l}{\textit{Hyperonic DDB$\Lambda\Xi^-$}} \\
    A1 base                         & $-28.726 \pm 0.083$ & $-28.759 \pm 0.126$ & $0.22$ \\
    $K_0=200$                       & $-29.121 \pm 0.094$ & $-29.394 \pm 0.358$ & $0.74$ \\
    $K_0=260$                       & $-28.608 \pm 0.075$ & $-28.681 \pm 0.225$ & $0.30$ \\
    $J_{\mathrm{sym}}=29$          & $-28.692 \pm 0.073$ & $-28.553 \pm 0.120$ & $0.99$ \\
    $J_{\mathrm{sym}}=36$          & $-28.923 \pm 0.111$ & $-29.016 \pm 0.226$ & $0.37$ \\
    PSR~J0614$-$3329                & $-31.439 \pm 0.197$ & $-31.533 \pm 0.259$ & $0.29$ \\
    PSR~J1231$-$1411                & $-29.117 \pm 0.122$ & $-29.240 \pm 0.195$ & $0.53$ \\
    PSR~J1614$-$2230                & $-28.744 \pm 0.117$ & $-28.789 \pm 0.109$ & $0.28$ \\
    \bottomrule
  \end{tabular*}
  \end{minipage}
\end{resulttablewide}

Table~\ref{tab:en-ultranest} compares the
EN with independent UltraNest calculations for both the nucleonic DDB and
hyperonic DDB$\Lambda\Xi^-$ models. Each EN entry is the equal-weight
evidence-space mean of three independently trained, frozen networks, with the
quoted uncertainty defined in Sec.~\ref{sec:method:en}. Where several
independent UltraNest seed runs are available, the UltraNest entry is their
inverse-variance combination with a scatter-aware uncertainty. For nucleonic
DDB, this applies to $K_0=260$ and $J_{\mathrm{sym}}=29$~MeV (two runs) and
$J_{\mathrm{sym}}=36$~MeV (three runs). For the nucleonic fiducial A1
configuration, the EN gives $\log\mathcal{Z}=-25.613\pm0.050$, consistent with
the UltraNest value $-25.467\pm0.242$ at $0.59\sigma$. With the astrophysical
likelihood fixed, the same frozen networks evaluate the four shifted nuclear
measurements with separations of only $0.08$--$0.33\sigma$ from UltraNest. The
three unseen-source configurations give separations of $0.85\sigma$ for
PSR~J0614$-$3329, $0.51\sigma$ for PSR~J1231$-$1411, and $0.06\sigma$ for
PSR~J1614$-$2230. All eight nucleonic differences are below $1\sigma$.

For hyperonic DDB$\Lambda\Xi^-$, the UltraNest entries
for A1, $J_{\mathrm{sym}}=29$~MeV, and PSR~J1614$-$2230 each combine two
independent seed runs in the same way. The fiducial A1 estimate is
$\log\mathcal{Z}=-28.726\pm0.083$, compared with $-28.759\pm0.126$ from
UltraNest, a separation of $0.22\sigma$. Relative to the corresponding
nucleonic A1 values, the EN and UltraNest imply Bayes factors of approximately
$22{:}1$ and $27{:}1$, respectively, in favor of the nucleonic composition. The
four shifted nuclear configurations agree within $0.30$--$0.99\sigma$. Both
methods give $K_0=260$~MeV a higher evidence than the base configuration and
$K_0=200$~MeV a lower one, consistent with a stiffer nucleonic sector
compensating hyperonic softening at high density. For the unseen
PSR~J0614$-$3329, PSR~J1231$-$1411, and PSR~J1614$-$2230 configurations, the EN
lies only $0.29\sigma$, $0.53\sigma$, and $0.28\sigma$ from UltraNest,
respectively.

Across all sixteen comparisons, evidence estimates from the EN agree with
UltraNest within the combined $1\sigma$ uncertainty (maximum $0.99\sigma$;
Table~\ref{tab:en-ultranest}). These estimates require no classical sampler;
UltraNest serves only as an independent reference. A stricter comparison with
independent A-NET importance-sampling evidences places every difference within
the EN uncertainty alone (maximum $0.96\sigma$; Table~\ref{tab:en-is}). Once
trained and frozen, the networks evaluate new covered configurations through
the two-dimensional integral in Eq.~\eqref{eq:en-green} for new NICER sources or
one evaluation of the second output for new nuclear data, avoiding further
nested-sampling runs. This amortizes repeated EOS evidence calculations and, to
our knowledge, is the first application of the Evidence-Network approach to
neutron-star EOS inference.

\raggedbottom
\subsection{Computational cost}
\label{sec:results:computational-cost}

Table~\ref{tab:cost} quantifies the computational advantage of
amortizing over the nuclear and NICER inputs. Once A-NET has been trained for
a given EOS composition, the frozen proposal analyzes a new configuration with
only the per-configuration full-likelihood correction, provided that the
predeclared importance-sampling gates are passed. For nucleonic DDB, each
A-NET+IS analysis takes about $1.7$--$13$~min with $8{,}000$ or $32{,}000$
likelihood evaluations, whereas UltraNest takes about $2.8$--$6.9$~h with
$3.5$--$8.7\times10^{5}$ evaluations, giving speedup factors of about
$25$--$240$. For hyperonic DDB$\Lambda\Xi^-$, A-NET+IS takes about
$2.9$--$8.5$~min with $25{,}000$ or $50{,}000$ likelihood evaluations, compared
with about $2.2$--$5.7$~h for UltraNest, giving speedup factors of about
$15$--$120$. All UltraNest runs use $1{,}000$ live points, except those for
PSR~J1614$-$2230, which use $2{,}000$. For nucleonic DDB, the one-time cost
before the first query is about $1.4$~h; including it, all eight
configurations take about $2.1$~h with A-NET+IS, compared with $33$--$35$~h
with UltraNest, that is, about $16$--$17$ times faster. For hyperonic
DDB$\Lambda\Xi^-$, the one-time cost is about $6.7$~h; including it, all eight
configurations take about $7.3$~h, compared with about $25$~h with UltraNest,
that is, about $3.4$ times faster. Together with the posterior agreement
demonstrated above, these comparisons show that A-NET can incorporate changed
nuclear measurements and previously unseen NICER sources within its training
coverage without retraining or loss of inference accuracy.

\begin{resulttablewide}
  \begin{minipage}{\textwidth}
  \caption{Computational cost of posterior inference for the nucleonic DDB and hyperonic DDB$\Lambda\Xi^-$ models: A-NET+IS for eight data configurations (A1, four nuclear-input changes, and three held-out NICER sources), TSNPE+MIS for A1, and UltraNest.}
  \label{tab:cost}
  \centering
  \scriptsize
  \setlength{\tabcolsep}{2pt}
  \begin{tabular*}{\textwidth}{@{\extracolsep{\fill}}p{0.1392\textwidth}p{0.0647\textwidth}p{0.0784\textwidth}p{0.1078\textwidth}p{0.0769\textwidth}p{0.1339\textwidth}p{0.1888\textwidth}p{0.0824\textwidth}@{}}
    \toprule
    & \multicolumn{3}{c}{A-NET+IS} & \multicolumn{3}{c}{UltraNest} & \tabularnewline
    \cmidrule(lr){2-4}\cmidrule(lr){5-7}
    \raggedright Configuration & \raggedright Time & \raggedright Likelihood\\evaluations & \raggedright ESS\\(efficiency) & \raggedright Time & \raggedright Likelihood\\evaluations & \raggedright ESS\\(efficiency) & \raggedright A-NET+IS\\faster by \tabularnewline
    \midrule
    \multicolumn{8}{@{}l}{\textit{Nucleonic DDB}} \tabularnewline
    \raggedright A1 (published data) & \raggedright $\approx$1.7~min & \raggedright 8,000 & \raggedright $\approx$530 (6.6\%) & \raggedright $\approx$3.3~h & \raggedright $\approx$4.1$\times10^{5}$ & \raggedright $\approx$4,500 (1.1\%) & \raggedright $\approx$120$\times$ \tabularnewline
    \raggedright $K_0=200$~MeV & \raggedright $\approx$5.2~min & \raggedright 8,000 & \raggedright $\approx$400 (4.9\%) & \raggedright $\approx$2.8~h & \raggedright $\approx$3.6$\times10^{5}$ & \raggedright $\approx$4,600 (1.3\%) & \raggedright $\approx$32$\times$ \tabularnewline
    \raggedright $K_0=260$~MeV & \raggedright $\approx$6.6~min & \raggedright 32,000 & \raggedright $\approx$990 (3.1\%) & \raggedright $\approx$4.5; 3.8~h & \raggedright $\approx$5.7$\times10^{5}$; 4.8$\times10^{5}$ & \raggedright $\approx$4,700; 4,700 (0.8\%; 1.0\%) & \raggedright $\approx$34--41$\times$ \tabularnewline
    \raggedright $J_{\rm sym}=29$~MeV & \raggedright $\approx$2.6~min & \raggedright 8,000 & \raggedright $\approx$1,300 (16.0\%) & \raggedright $\approx$2.8; 3.0~h & \raggedright $\approx$3.5$\times10^{5}$; 3.8$\times10^{5}$ & \raggedright $\approx$4,800; 4,800 (1.3\%; 1.3\%) & \raggedright $\approx$66--70$\times$ \tabularnewline
    \raggedright $J_{\rm sym}=36$~MeV & \raggedright $\approx$9.1~min & \raggedright 32,000 & \raggedright $\approx$760 (2.4\%) & \raggedright $\approx$3.8; 4.1; 4.4~h & \raggedright $\approx$4.8$\times10^{5}$; 5.2$\times10^{5}$; 5.6$\times10^{5}$ & \raggedright $\approx$4,500; 4,600; 4,700 (0.9\%; 0.9\%; 0.8\%) & \raggedright $\approx$25--29$\times$ \tabularnewline
    \raggedright PSR~J0614$-$3329 & \raggedright $\approx$1.7~min & \raggedright 8,000 & \raggedright $\approx$450 (5.6\%) & \raggedright $\approx$6.9~h & \raggedright $\approx$8.7$\times10^{5}$ & \raggedright $\approx$4,500 (0.5\%) & \raggedright $\approx$240$\times$ \tabularnewline
    \raggedright PSR~J1231$-$1411 & \raggedright $\approx$1.8~min & \raggedright 8,000 & \raggedright $\approx$1,500 (19.4\%) & \raggedright $\approx$3.2~h & \raggedright $\approx$4.1$\times10^{5}$ & \raggedright $\approx$4,500 (1.1\%) & \raggedright $\approx$110$\times$ \tabularnewline
    \raggedright PSR~J1614$-$2230 & \raggedright $\approx$13~min & \raggedright 32,000 & \raggedright $\approx$3,300 (10.3\%) & \raggedright $\approx$6.6~h & \raggedright $\approx$8.4$\times10^{5}$ & \raggedright $\approx$8,700 (1.0\%) & \raggedright $\approx$30$\times$ \tabularnewline
    \addlinespace[2pt]
    \multicolumn{8}{@{}l}{\textit{Hyperonic DDB$\Lambda\Xi^-$}} \tabularnewline
    \raggedright A1 (published data) & \raggedright $\approx$4.2~min & \raggedright 25,000 & \raggedright $\approx$5,600 (22.3\%) & \raggedright $\approx$2.4; 2.5~h & \raggedright $\approx$3.7$\times10^{5}$; 3.9$\times10^{5}$ & \raggedright $\approx$4,300; 4,500 (1.2\%; 1.1\%) & \raggedright $\approx$34--36$\times$ \tabularnewline
    \raggedright $K_0=200$~MeV & \raggedright $\approx$4.2~min & \raggedright 25,000 & \raggedright $\approx$3,700 (14.9\%) & \raggedright $\approx$2.2~h & \raggedright $\approx$3.4$\times10^{5}$ & \raggedright $\approx$4,500 (1.3\%) & \raggedright $\approx$31$\times$ \tabularnewline
    \raggedright $K_0=260$~MeV & \raggedright $\approx$4.3~min & \raggedright 25,000 & \raggedright $\approx$4,300 (17.2\%) & \raggedright $\approx$2.6~h & \raggedright $\approx$4.0$\times10^{5}$ & \raggedright $\approx$4,200 (1.0\%) & \raggedright $\approx$36$\times$ \tabularnewline
    \raggedright $J_{\rm sym}=29$~MeV & \raggedright $\approx$8.5~min & \raggedright 50,000 & \raggedright $\approx$6,400 (12.8\%) & \raggedright $\approx$2.2; 2.2~h & \raggedright $\approx$3.4$\times10^{5}$; 3.4$\times10^{5}$ & \raggedright $\approx$4,400; 4,400 (1.3\%; 1.3\%) & \raggedright $\approx$15--16$\times$ \tabularnewline
    \raggedright $J_{\rm sym}=36$~MeV & \raggedright $\approx$4.3~min & \raggedright 25,000 & \raggedright $\approx$3,800 (15.1\%) & \raggedright $\approx$2.7~h & \raggedright $\approx$4.2$\times10^{5}$ & \raggedright $\approx$4,700 (1.1\%) & \raggedright $\approx$39$\times$ \tabularnewline
    \raggedright PSR~J0614$-$3329 & \raggedright $\approx$4.2~min & \raggedright 25,000 & \raggedright $\approx$2,200 (9.0\%) & \raggedright $\approx$3.1~h & \raggedright $\approx$4.7$\times10^{5}$ & \raggedright $\approx$4,500 (1.0\%) & \raggedright $\approx$43$\times$ \tabularnewline
    \raggedright PSR~J1231$-$1411 & \raggedright $\approx$7.2~min & \raggedright 50,000 & \raggedright $\approx$6,700 (13.4\%) & \raggedright $\approx$4.1~h & \raggedright $\approx$6.4$\times10^{5}$ & \raggedright $\approx$4,400 (0.7\%) & \raggedright $\approx$34$\times$ \tabularnewline
    \raggedright PSR~J1614$-$2230 & \raggedright $\approx$2.9~min & \raggedright 25,000 & \raggedright $\approx$5,200 (20.9\%) & \raggedright $\approx$5.5; 5.7~h & \raggedright $\approx$8.4$\times10^{5}$; 8.7$\times10^{5}$ & \raggedright $\approx$8,200; 8,600 (1.0\%; 1.0\%) & \raggedright $\approx$110--120$\times$ \tabularnewline
    \bottomrule
  \end{tabular*}
  \par\vspace{3pt}
  \begin{tabular*}{\textwidth}{@{\extracolsep{\fill}}p{0.4627\textwidth}p{0.1961\textwidth}p{0.1912\textwidth}p{0.1176\textwidth}@{}}
    \toprule
    \raggedright Summary & \raggedright Time & \raggedright Likelihood evaluations & \raggedright Compared with UltraNest \tabularnewline
    \midrule
    \multicolumn{4}{@{}l}{\textit{Nucleonic DDB}} \tabularnewline
    \raggedright A-NET+IS one-time cost before the first query & \raggedright $\approx$1.4~h & \raggedright 0 & \raggedright -- \tabularnewline
    \raggedright TSNPE+MIS, A1 only (ESS $\approx$31,000, 13.0\%) & \raggedright $\approx$1.4~h (reference-equivalent estimate) & \raggedright $\approx$2.4$\times10^{5}$ & \raggedright $\approx$2.3$\times$ faster \tabularnewline
    \raggedright All eight configurations, including the one-time cost; one UltraNest run per configuration (range over independent seeds) & \raggedright $\approx$2.1~h vs $\approx$33--35~h & \raggedright $\approx$1.4$\times10^{5}$ vs $\approx$4.2--4.4$\times10^{6}$ & \raggedright $\approx$16--17$\times$ faster \tabularnewline
    \addlinespace[2pt]
    \multicolumn{4}{@{}l}{\textit{Hyperonic DDB$\Lambda\Xi^-$}} \tabularnewline
    \raggedright A-NET+IS one-time cost before the first query & \raggedright $\approx$6.7~h & \raggedright $\approx$2.9$\times10^{6}$ & \raggedright -- \tabularnewline
    \raggedright TSNPE+MIS, A1 only (ESS $\approx$1,500, 0.2\%) & \raggedright $\approx$1.5~h (recorded-stage parallel-equivalent) & \raggedright $\approx$7.8$\times10^{5}$ & \raggedright $\approx$1.6$\times$ faster \tabularnewline
    \raggedright All eight configurations, including the one-time cost; one UltraNest run per configuration (range over independent seeds) & \raggedright $\approx$7.3~h vs $\approx$25~h & \raggedright $\approx$3.2$\times10^{6}$ vs $\approx$3.8--3.9$\times10^{6}$ & \raggedright $\approx$3.4$\times$ faster \tabularnewline
    \bottomrule
  \end{tabular*}
  \end{minipage}
\end{resulttablewide}

\begin{resulttablewide}
  \begin{minipage}{\textwidth}
  \caption{Computational cost of Bayesian-evidence estimation with the Green-function Evidence Network (EN), compared with UltraNest, for the nucleonic DDB and hyperonic DDB$\Lambda\Xi^-$ models.}
  \label{tab:en-cost}
  \centering
  \scriptsize
  \setlength{\tabcolsep}{1.5pt}
  \begin{tabular*}{\textwidth}{@{\extracolsep{\fill}}p{0.1392\textwidth}p{0.1235\textwidth}p{0.0784\textwidth}p{0.0784\textwidth}p{0.0769\textwidth}p{0.1339\textwidth}p{0.1888\textwidth}p{0.1357\textwidth}@{}}
    \toprule
    & \multicolumn{3}{c}{EN} & \multicolumn{3}{c}{UltraNest} & \tabularnewline
    \cmidrule(lr){2-4}\cmidrule(lr){5-7}
    \raggedright Configuration & \raggedright Time & \raggedright Likelihood\\evaluations & \raggedright ESS\\(efficiency) & \raggedright Time & \raggedright Likelihood\\evaluations & \raggedright ESS\\(efficiency) & \raggedright EN\\faster by \tabularnewline
    \midrule
    \multicolumn{8}{@{}l}{\textit{Nucleonic DDB}} \tabularnewline
    \raggedright A1 (published data) & \raggedright $\approx$0.026~s (all five) & \raggedright 0 & \raggedright -- & \raggedright $\approx$3.3~h & \raggedright $\approx$4.1$\times10^{5}$ & \raggedright $\approx$4,500 (1.1\%) & \raggedright $\approx$450,000$\times$ \tabularnewline
    \raggedright $K_0=200$~MeV & \raggedright $\approx$0.026~s (all five) & \raggedright 0 & \raggedright -- & \raggedright $\approx$2.8~h & \raggedright $\approx$3.6$\times10^{5}$ & \raggedright $\approx$4,600 (1.3\%) & \raggedright $\approx$390,000$\times$ \tabularnewline
    \raggedright $K_0=260$~MeV & \raggedright $\approx$0.026~s (all five) & \raggedright 0 & \raggedright -- & \raggedright $\approx$4.5; 3.8~h & \raggedright $\approx$5.7$\times10^{5}$; 4.8$\times10^{5}$ & \raggedright $\approx$4,700; 4,700 (0.8\%; 1.0\%) & \raggedright $\approx$520,000--620,000$\times$ \tabularnewline
    \raggedright $J_{\rm sym}=29$~MeV & \raggedright $\approx$0.026~s (all five) & \raggedright 0 & \raggedright -- & \raggedright $\approx$2.8; 3.0~h & \raggedright $\approx$3.5$\times10^{5}$; 3.8$\times10^{5}$ & \raggedright $\approx$4,800; 4,800 (1.3\%; 1.3\%) & \raggedright $\approx$390,000--410,000$\times$ \tabularnewline
    \raggedright $J_{\rm sym}=36$~MeV & \raggedright $\approx$0.026~s (all five) & \raggedright 0 & \raggedright -- & \raggedright $\approx$3.8; 4.1; 4.4~h & \raggedright $\approx$4.8$\times10^{5}$; 5.2$\times10^{5}$; 5.6$\times10^{5}$ & \raggedright $\approx$4,500; 4,600; 4,700 (0.9\%; 0.9\%; 0.8\%) & \raggedright $\approx$530,000--610,000$\times$ \tabularnewline
    \raggedright PSR~J0614$-$3329 & \raggedright $\approx$3.4~s & \raggedright 0 & \raggedright -- & \raggedright $\approx$6.9~h & \raggedright $\approx$8.7$\times10^{5}$ & \raggedright $\approx$4,500 (0.5\%) & \raggedright $\approx$7,400$\times$ \tabularnewline
    \raggedright PSR~J1231$-$1411 & \raggedright $\approx$3.5~s & \raggedright 0 & \raggedright -- & \raggedright $\approx$3.2~h & \raggedright $\approx$4.1$\times10^{5}$ & \raggedright $\approx$4,500 (1.1\%) & \raggedright $\approx$3,300$\times$ \tabularnewline
    \raggedright PSR~J1614$-$2230 & \raggedright $\approx$3.5~s & \raggedright 0 & \raggedright -- & \raggedright $\approx$6.6~h & \raggedright $\approx$8.4$\times10^{5}$ & \raggedright $\approx$8,700 (1.0\%) & \raggedright $\approx$6,700$\times$ \tabularnewline
    \addlinespace[2pt]
    \multicolumn{8}{@{}l}{\textit{Hyperonic DDB$\Lambda\Xi^-$}} \tabularnewline
    \raggedright A1 (published data) & \raggedright $\approx$0.027~s (all five) & \raggedright 0 & \raggedright -- & \raggedright $\approx$2.4; 2.5~h & \raggedright $\approx$3.7$\times10^{5}$; 3.9$\times10^{5}$ & \raggedright $\approx$4,300; 4,500 (1.2\%; 1.1\%) & \raggedright $\approx$320,000--330,000$\times$ \tabularnewline
    \raggedright $K_0=200$~MeV & \raggedright $\approx$0.027~s (all five) & \raggedright 0 & \raggedright -- & \raggedright $\approx$2.2~h & \raggedright $\approx$3.4$\times10^{5}$ & \raggedright $\approx$4,500 (1.3\%) & \raggedright $\approx$290,000$\times$ \tabularnewline
    \raggedright $K_0=260$~MeV & \raggedright $\approx$0.027~s (all five) & \raggedright 0 & \raggedright -- & \raggedright $\approx$2.6~h & \raggedright $\approx$4.0$\times10^{5}$ & \raggedright $\approx$4,200 (1.0\%) & \raggedright $\approx$340,000$\times$ \tabularnewline
    \raggedright $J_{\rm sym}=29$~MeV & \raggedright $\approx$0.027~s (all five) & \raggedright 0 & \raggedright -- & \raggedright $\approx$2.2; 2.2~h & \raggedright $\approx$3.4$\times10^{5}$; 3.4$\times10^{5}$ & \raggedright $\approx$4,400; 4,400 (1.3\%; 1.3\%) & \raggedright $\approx$290,000$\times$ \tabularnewline
    \raggedright $J_{\rm sym}=36$~MeV & \raggedright $\approx$0.027~s (all five) & \raggedright 0 & \raggedright -- & \raggedright $\approx$2.7~h & \raggedright $\approx$4.2$\times10^{5}$ & \raggedright $\approx$4,700 (1.1\%) & \raggedright $\approx$360,000$\times$ \tabularnewline
    \raggedright PSR~J0614$-$3329 & \raggedright $\approx$3.4~s & \raggedright 0 & \raggedright -- & \raggedright $\approx$3.1~h & \raggedright $\approx$4.7$\times10^{5}$ & \raggedright $\approx$4,500 (1.0\%) & \raggedright $\approx$3,300$\times$ \tabularnewline
    \raggedright PSR~J1231$-$1411 & \raggedright $\approx$3.5~s & \raggedright 0 & \raggedright -- & \raggedright $\approx$4.1~h & \raggedright $\approx$6.4$\times10^{5}$ & \raggedright $\approx$4,400 (0.7\%) & \raggedright $\approx$4,200$\times$ \tabularnewline
    \raggedright PSR~J1614$-$2230 & \raggedright $\approx$3.5~s & \raggedright 0 & \raggedright -- & \raggedright $\approx$5.5; 5.7~h & \raggedright $\approx$8.4$\times10^{5}$; 8.7$\times10^{5}$ & \raggedright $\approx$8,200; 8,600 (1.0\%; 1.0\%) & \raggedright $\approx$5,600--5,800$\times$ \tabularnewline
    \bottomrule
  \end{tabular*}
  \par\vspace{3pt}
  \begin{tabular*}{\textwidth}{@{\extracolsep{\fill}}p{0.4627\textwidth}p{0.1961\textwidth}p{0.1912\textwidth}p{0.1176\textwidth}@{}}
    \toprule
    \raggedright Summary & \raggedright Time & \raggedright Likelihood evaluations & \raggedright Compared with UltraNest \tabularnewline
    \midrule
    \multicolumn{4}{@{}l}{\textit{Nucleonic DDB}} \tabularnewline
    \raggedright EN one-time cost before the first query & \raggedright $\approx$30~min & \raggedright $\approx$6.2$\times10^{4}$ & \raggedright -- \tabularnewline
    \raggedright All eight configurations, including the one-time cost; one UltraNest run per configuration (range over independent seeds) & \raggedright $\approx$31~min vs $\approx$33--35~h & \raggedright $\approx$6.2$\times10^{4}$ vs $\approx$4.2--4.4$\times10^{6}$ & \raggedright $\approx$65--68$\times$ faster \tabularnewline
    \addlinespace[2pt]
    \multicolumn{4}{@{}l}{\textit{Hyperonic DDB$\Lambda\Xi^-$}} \tabularnewline
    \raggedright EN one-time cost before the first query & \raggedright $\approx$55~min & \raggedright $\approx$1.5$\times10^{5}$ & \raggedright -- \tabularnewline
    \raggedright All eight configurations, including the one-time cost; one UltraNest run per configuration (range over independent seeds) & \raggedright $\approx$55~min vs $\approx$25~h & \raggedright $\approx$1.5$\times10^{5}$ vs $\approx$3.8--3.9$\times10^{6}$ & \raggedright $\approx$27$\times$ faster \tabularnewline
    \bottomrule
  \end{tabular*}
  \end{minipage}
\end{resulttablewide}

We time only successful production runs. When a stage runs in parallel, we count
only its longest part. UltraNest times are equivalent times,
$t_{\rm eq}=N_{\rm eval}/q_{\rm ref}$. The reference rate $q_{\rm ref}$ is $35.0$
likelihood evaluations per second for the nucleonic model and $42.9$ for the
hyperonic model, with uncertainties of $15\%$ and $10\%$. The one-time A-NET cost
includes the preparation of the training data and the network training. For the
hyperonic model, it also includes building the defensive proposal and uses
$2.9\times10^{6}$ likelihood evaluations in total. A-NET, TSNPE, and EN all use a
shared EOS/TOV bank. The cost of building this bank is not included in any of the
times. Table~\ref{tab:cost} also does not include the separate calculation of the
nucleonic A1 mass--tidal band (Sec.~\ref{sec:method:anet}). The stellar curves of
its $3.2\times10^{5}$ new draws took about $18$~min. In Tables~\ref{tab:cost}
and~\ref{tab:en-cost}, semicolons separate independent UltraNest runs. For these
entries, the speedup is given as a range over the runs. Times, counts, ESS values,
and speedups are rounded to two significant figures. The complete timing ledger is
provided with the online package.

TSNPE+MIS provides a complementary, target-specific
neural analysis. Unlike A-NET, its simulations and training must be repeated
for each observation. For nucleonic A1, TSNPE+MIS takes about $1.4$~h with
about $2.4\times10^{5}$ likelihood evaluations, compared with about $3.3$~h and
$4.1\times10^{5}$ evaluations for UltraNest, that is, about $2.3$ times faster.
For hyperonic A1, it takes about $1.5$~h, compared with about $2.4$--$2.5$~h
for UltraNest, that is, about $1.6$ times faster, although it uses about
$7.8\times10^{5}$ likelihood evaluations, about twice the UltraNest count of
$3.7$--$3.9\times10^{5}$. To compare TSNPE+MIS and UltraNest on the same footing, the nucleonic
TSNPE+MIS time is expressed at the UltraNest reference rate, and the hyperonic time
is the sum of the recorded stage times, with the EOS-solving time evaluated for the
production EOS solver. In Table~\ref{tab:cost}, the ESS efficiency is the
ESS divided by the number of likelihood evaluations.

Table~\ref{tab:cost} compares A-NET+IS with a new UltraNest run for each
configuration. Two further comparisons put this speedup in context. First, the two
methods return different numbers of effective samples. Per effective sample,
A-NET+IS is about $2.8$--$38$ times faster than UltraNest for nucleonic DDB and
$22$--$73$ times faster for hyperonic DDB$\Lambda\Xi^-$. Second, instead of a new
run, one could reweight the stored A1 UltraNest posterior to the new data by the
likelihood ratio. This costs only one likelihood evaluation per stored sample, but
it is not reliable. For nucleonic DDB, the reweighted median of $R_{1.4}$ lies
below the fresh UltraNest result in all seven changed configurations, by
$0.013$--$0.056$~km, which is up to $3.3$ times the Monte Carlo error. Its 5th and
95th percentiles deviate by up to $0.083$~km. For hyperonic DDB$\Lambda\Xi^-$,
reweighting to the $J_{\rm sym}$ shifts retains an ESS of only about $580$, and the
5th percentile for $J_{\rm sym}=36$~MeV deviates by $0.062$~km. In all these
configurations, A-NET+IS agrees with the fresh UltraNest runs within $0.014$~km in
the median and about $0.04$~km in the 5th and 95th percentiles. Reweighting reuses
the stored samples and inherits their sampling errors, whereas A-NET+IS draws new
samples from a proposal adapted to each configuration.

Table~\ref{tab:en-cost} shows the corresponding advantage for
Bayesian-evidence calculations. After a one-time cost before the first query of
about $30$~min for nucleonic and $55$~min for hyperonic DDB, the EN returns the
evidences of all five nuclear configurations together in about $0.03$~s and
the evidence for each held-out NICER source in about $3.5$~s, without further
likelihood evaluations. The one-time EN cost consists of EN-specific data preparation (about $12$ and
$47$~min for the nucleonic and hyperonic models) and ensemble training (about $18$
and $8$~min). These are warm query times. The joint query time includes loading
the three frozen networks, and each NICER query time includes building the source
density table once. The joint query time is counted in full for each of the five
nuclear configurations. No ESS is given for the EN, because a query returns the
evidence without samples. For a new NICER source, the quoted uncertainty also
needs a bank estimate. This estimate evaluates the new source term on the stored
bank curves, without new EOS or TOV solutions, and takes a further $4$--$9$~s.
The matched UltraNest calculations take about
$2.8$--$6.9$~h for nucleonic and $2.2$--$5.7$~h for hyperonic DDB. The
per-configuration speedup factors are therefore about $290{,}000$--$620{,}000$
for the nuclear configurations and $3{,}300$--$7{,}400$ for the held-out
sources. For nucleonic DDB, including the one-time cost, all eight
configurations take about $31$~min with the EN, compared with $33$--$35$~h with
UltraNest, that is, about $65$--$68$ times faster. For hyperonic
DDB$\Lambda\Xi^-$, they take about $55$~min, compared with about $25$~h with
UltraNest, that is, about $27$ times faster. Because every EN--UltraNest
difference in Table~\ref{tab:en-ultranest} is below $1\sigma$, this speedup is
obtained without degrading the accuracy of the evidence. The UltraNest entries
in Tables~\ref{tab:cost} and \ref{tab:en-cost} are the same calculations,
because nested sampling returns the posterior and evidence together.

\subsection{Computing hardware}
\label{sec:results:hardware}
All computations ran on two desktop workstations, on 64-core nodes of a CPU
cluster, and on rented RunPod cloud instances. Each desktop has an Intel Core
i7-13700F CPU (24 threads) and an NVIDIA RTX~3060~Ti GPU. The nucleonic A-NET
network was trained, and the nucleonic A-NET+IS analyses were run, on one desktop
with its RTX~3060~Ti GPU and up to $14$ CPU workers. The nucleonic TSNPE+MIS
analysis ran on the second desktop. For hyperonic DDB$\Lambda\Xi^-$, the A-NET
training data and the nine-dimensional head used an NVIDIA H100 NVL GPU, the
seven-dimensional head an NVIDIA RTX~4090 GPU, and the A-NET proposals an NVIDIA
H100 80\,GB GPU, all on RunPod. The hyperonic TSNPE flows were fitted on RunPod NVIDIA
RTX~4090 GPUs. The EN training labels and networks used RunPod NVIDIA A40 and
RTX~4090 GPUs. The EN queries ran on one desktop RTX~3060~Ti GPU, and the bank
estimate for a new NICER source on one desktop CPU. The exact likelihood
evaluations ran on the desktops and the cluster nodes, with part of the hyperonic
A-NET preparation on a RunPod CPU instance. The UltraNest reference rates were
measured with $20$ likelihood processes on a desktop workstation in three complete
runs, and with $56$ processes on one cluster node in four complete runs. The
stellar curves of the nucleonic A1 mass--tidal calculation used $56$ parallel
workers. The machine and the number of CPU cores used for each stage depended on
what was available at the time, since the computing resources were limited. The
hardware therefore differs between stages. Absolute times depend on the hardware,
whereas the numbers of likelihood evaluations and the sample efficiencies do not.

\flushbottom

\section{Conclusions and outlook}
\label{sec:conclusions}

We have developed two complementary, likelihood-corrected
neural routes for microscopic neutron-star EOS inference. TSNPE+MIS performs
target-specific sequential inference by concentrating new simulations around a
single observation, whereas A-NET+IS amortizes the mapping from changing
nuclear and NICER inputs to the EOS-parameter posterior and can be reused
across covered data configurations. In both approaches, the neural density is
used as an IS proposal rather than reported directly: weights
calculated from the original prior and complete multimessenger likelihood
correct residual neural approximation error, while the ESS
monitors the quality of each correction. Applied to nucleonic DDB
and hyperonic DDB$\Lambda\Xi^-$ matter, both routes closely agree with
UltraNest in the inferred posterior structure and 90\% mass--radius and
mass--tidal-deformability bands. The mean band-edge
differences are $0.024$--$0.033$~km and $1.02$--$1.03\%$ for TSNPE+MIS, and
$0.019$--$0.050$~km and $0.86$--$1.13\%$ for A-NET+IS. This agreement
establishes two accurate but operationally distinct
neural inference strategies.

The two routes differ in how their cost
scales when the input data change. TSNPE+MIS concentrates its simulations on
one specified target, providing an efficient neural route for a focused
analysis. A-NET+IS instead changes the scaling of an inference campaign by
sharing one training cost across later queries and reducing calculations that
require hours with UltraNest to only minutes. The frozen A-NET does more than
reproduce the fiducial A1 posterior: it follows the distinct shifts produced by
four nuclear-input changes and three NICER sources excluded from training,
including the lower-radius preference introduced by the new PSR~J1614$-$2230
measurement, for both nucleonic and hyperonic matter. The
EN extends this reuse from parameter inference to model
selection, turning each evidence calculation into a network
query of about $0.03$~s for new nuclear data and about $3.5$~s for a new NICER
source, rather than a new nested-sampling calculation. All sixteen EN evidences
agree with UltraNest within $1\sigma$ and with independent importance-sampling evidences within the EN uncertainties. Under
the adopted likelihood, the two methods give Bayes factors of about $22$ (EN) and $27$ (UltraNest) in
favor of nucleonic DDB over the particular hyperonic extension. The speedup comes from reusing trained
networks, not from approximating the likelihood.

These gains apply within the conditioning domain of the
frozen networks. For A-NET, this domain comprises the nuclear data, the three
NICER sources, and the GW and pQCD controls of Eq.~\eqref{eq:anet-context};
more general changes to the GW170817, maximum-mass, or pQCD terms, or to the
EOS family, require an expanded proposal or training domain whenever the
existing proposal does not provide sufficient support. Likewise, the inferred
composition preference
compares nucleonic DDB with the specific DDB$\Lambda\Xi^-$ model under the
adopted priors and should not be interpreted as model-independent evidence
against hyperonic matter. The source tests replace one NICER source at a time. They establish source
substitution within the tested domain, not the accumulation of many new
observations. Reweighting a stored posterior is cheaper, but in our tests it
shifted the median $R_{1.4}$ by up to $0.056$~km and the 5th and 95th percentiles
by up to $0.083$~km, whereas A-NET+IS agreed with fresh UltraNest runs within
$0.014$~km in the median. A natural extension for the general inference
package is a second permutation-invariant aggregation over the individual
source embeddings, trained across variable source counts, as introduced for
neural posterior estimation of model-agnostic EOS representations in
Ref.~\cite{Carvalho:2026nbl}. This would allow
users to supply arbitrary numbers of NICER and GW observations directly while
retaining full-likelihood correction, ESS diagnostics, and independent
sampler validation. This extension is in preparation.

\noindent\begin{minipage}{\columnwidth}
\section*{Data availability}
\vspace{-0.75em}

The analysis uses NumPy~\cite{Harris:2020numpy}, SciPy~\cite{Virtanen:2020scipy},
Numba~\cite{Lam:2015numba}, JAX~\cite{Bradbury:2018jax},
PyTorch~\cite{Paszke:2019pytorch}, zuko~\cite{Rozet:2022zuko},
sbi~\cite{Boelts:2025sbi}, scikit-learn~\cite{Pedregosa:2011sklearn}, and
UltraNest~\cite{Buchner:2021cql}. The figures were made with
Matplotlib~\cite{Hunter:2007matplotlib} and seaborn~\cite{Waskom:2021seaborn}.
The code of this work is released as the open-source package NeuronStar.
Source code, trained-network checkpoints, posterior and
evidence samples, numerical data, the timing values of
Tables~\ref{tab:cost} and~\ref{tab:en-cost}, reproduction scripts, and
environment files that pin the package versions are publicly available at
\url{https://github.com/prashantstar123/NeuronStar}.
\end{minipage}

\vspace{-0.5em}
\begin{acknowledgments}
\vspace{-0.75em}
P.~Thakur acknowledges support from the National Research
Foundation of Korea (NRF), funded by the Korean Ministry of Science and ICT
(MSIT) under Grant
No.~RS-2024-00457037, and from the 2025 Yonsei University Research Fund
(Yonsei University Frontier Fellowship for Postdoctoral Researchers).
\end{acknowledgments}

\appendix

\section{Nuclear-data amortization for hyperonic matter}
\label{app:hyperonic-nuclear-amortization}

Figure~\ref{fig:hyp-nuclear-amortization} and
Table~\ref{tab:hyp-nuclear-amortization-summary} repeat the nuclear-data
amortization test of Fig.~\ref{fig:nuclear-amortization} and
Table~\ref{tab:nuclear-amortization-summary} for the hyperonic
DDB$\Lambda\Xi^-$ model, using the same frozen hyperonic A-NET as in the
fiducial A1 analysis. Without retraining, A-NET+IS
reproduces the UltraNest 90\% mass--radius bands in all four configurations,
with mean absolute band-edge separations of $0.017$, $0.024$, $0.010$, and
$0.012$~km for $K_0=200$, $K_0=260$, $J_{\mathrm{sym}}=29$, and
$J_{\mathrm{sym}}=36$~MeV, respectively. Across all configurations, the largest
differences between the posterior medians are $0.003\,M_\odot$ in $M_{\max}$,
$0.003$~km in $R(M_{\max})$, and $0.013$~km in $R_{1.4}$, with overlapping
central 90\% intervals in every case. Increasing $K_0$ from $200$ to
$260$~MeV raises $R_{1.4}$ by about $0.19$~km with UltraNest and $0.20$~km
with A-NET+IS.

\begin{resultfigurewide}
  \centering
  \includegraphics[width=0.84\textwidth]{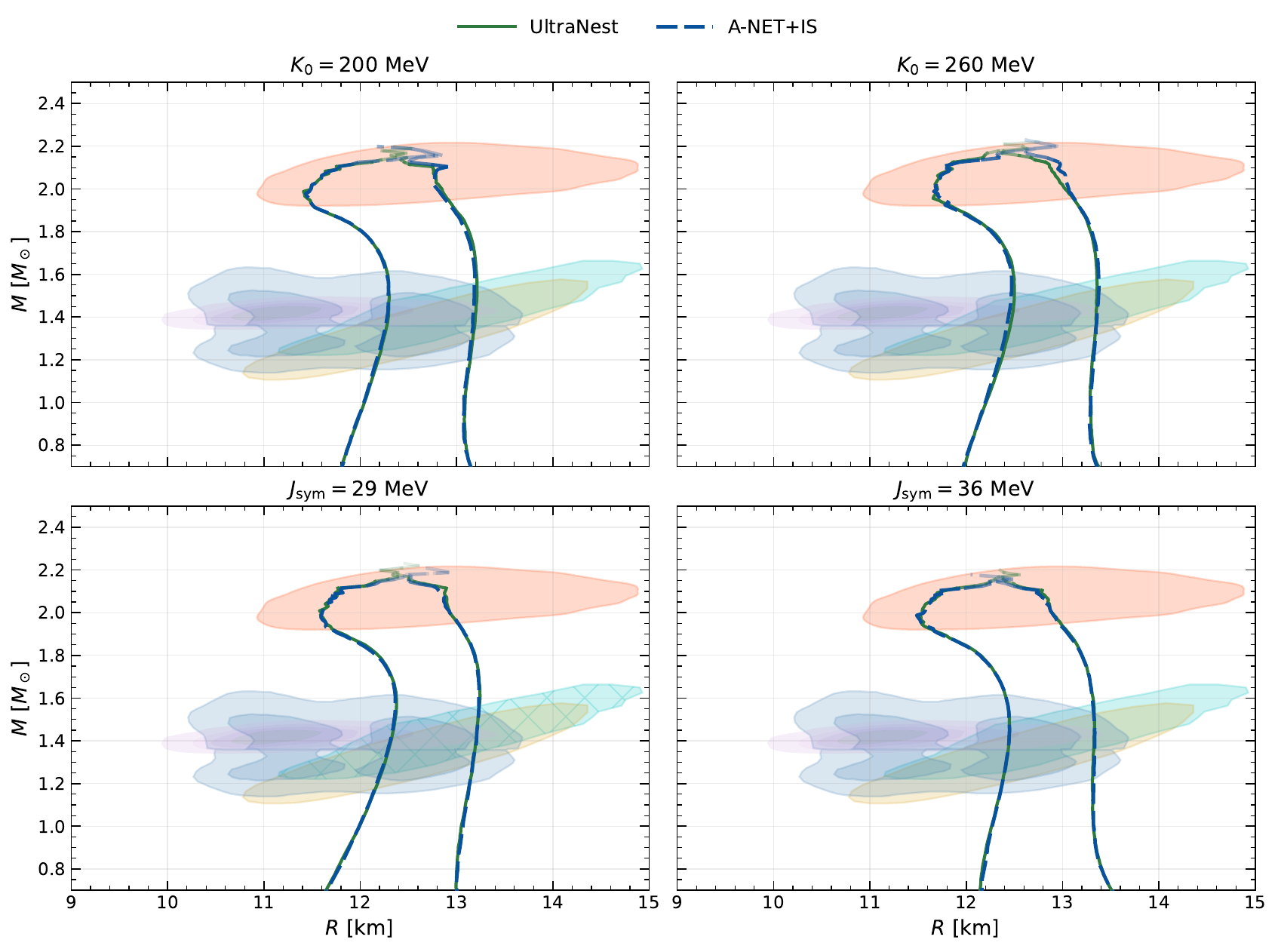}
  \caption{Hyperonic DDB$\Lambda\Xi^-$ version of
  Fig.~\ref{fig:nuclear-amortization}: 90\% credible mass--radius bands for
  the four shifted nuclear-data configurations obtained with UltraNest (solid
  green) and A-NET+IS (dashed blue). Observational
  constraints as in Fig.~\ref{fig:a1}.}
  \label{fig:hyp-nuclear-amortization}
  \par\vspace{4pt}
  \begin{minipage}{0.98\textwidth}
    \centering
    \makeatletter\def\@captype{table}\makeatother
    \caption{Hyperonic DDB$\Lambda\Xi^-$ version of
    Table~\ref{tab:nuclear-amortization-summary}: posterior medians and
    central 90\% credible intervals for the four shifted nuclear-data
    configurations. $M_{\max}$ is in $M_\odot$; $R(M_{\max})$ and $R_{1.4}$ are
    in km.}
    \label{tab:hyp-nuclear-amortization-summary}
    \footnotesize
    \setlength{\tabcolsep}{10pt}
    \renewcommand{\arraystretch}{1.18}
    \begin{tabular}{@{}llccc@{}}
      \toprule
      Configuration & Method & $M_{\max}$ & $R(M_{\max})$ & $R_{1.4}$ \\
      \midrule
      $K_0=200$ & UltraNest
        & $2.01^{+0.10}_{-0.09}$
        & $11.36^{+0.53}_{-0.47}$
        & $12.71^{+0.49}_{-0.42}$ \\
       & A-NET+IS
        & $2.01^{+0.10}_{-0.09}$
        & $11.36^{+0.51}_{-0.46}$
        & $12.72^{+0.45}_{-0.44}$ \\
      \addlinespace[2pt]
      $K_0=260$ & UltraNest
        & $2.03^{+0.09}_{-0.10}$
        & $11.60^{+0.50}_{-0.46}$
        & $12.90^{+0.44}_{-0.43}$ \\
       & A-NET+IS
        & $2.03^{+0.10}_{-0.10}$
        & $11.60^{+0.52}_{-0.49}$
        & $12.91^{+0.44}_{-0.47}$ \\
      \addlinespace[2pt]
      $J_{\mathrm{sym}}=29$ & UltraNest
        & $2.02^{+0.10}_{-0.10}$
        & $11.52^{+0.49}_{-0.45}$
        & $12.74^{+0.46}_{-0.42}$ \\
       & A-NET+IS
        & $2.02^{+0.10}_{-0.10}$
        & $11.52^{+0.51}_{-0.46}$
        & $12.75^{+0.46}_{-0.44}$ \\
      \addlinespace[2pt]
      $J_{\mathrm{sym}}=36$ & UltraNest
        & $2.02^{+0.09}_{-0.09}$
        & $11.44^{+0.52}_{-0.46}$
        & $12.87^{+0.46}_{-0.43}$ \\
       & A-NET+IS
        & $2.02^{+0.09}_{-0.10}$
        & $11.44^{+0.52}_{-0.47}$
        & $12.86^{+0.47}_{-0.42}$ \\
      \bottomrule
    \end{tabular}
  \end{minipage}
\end{resultfigurewide}

\section{Calibration of the A-NET proposals}
\label{app:anet-calibration}

Figure~\ref{fig:anet-pp} tests the marginal
calibration of the frozen A-NET proposals before importance sampling. The test
cases are generated with fresh random seeds in the same way as the training
data and were not used in training: $1{,}000$ cases for nucleonic DDB and $500$
for the nine-parameter network of the hyperonic DDB$\Lambda\Xi^-$ A-NET. For
each case, the frozen network is queried without retraining, and for every
parameter we record the rank of the true value, that is, the fraction of
proposal draws below it. The hyperonic network samples in bounded coordinates,
so all of its draws lie inside the prior box; the nucleonic network places
$3.2\%$ of its draws outside the box, and these draws, which receive zero
importance weight, do not enter its ranks. For a
calibrated proposal these
ranks are uniformly distributed: the true value falls below the $p$ quantile of
the proposal in a fraction $p$ of the cases, and each P--P curve follows the
diagonal within the grey binomial bands. For every parameter, the
Kolmogorov--Smirnov test is consistent with uniform ranks, with $p$ values of
at least $0.07$ (nucleonic) and $0.08$ (hyperonic). Because the reported
posteriors are importance-corrected with the full likelihood, this test
measures the quality of the proposals. Its cases reuse the EOS rows of the training bank with new
source configurations and nuclear-data noise. It therefore probes generalization
over data configurations rather than over EOS realizations. The independent test is the comparison of the final posteriors with UltraNest,
which uses no bank (Sec.~\ref{sec:results}).

\begin{resultfigurewide}
  \centering
  \includegraphics[width=0.75\textwidth]{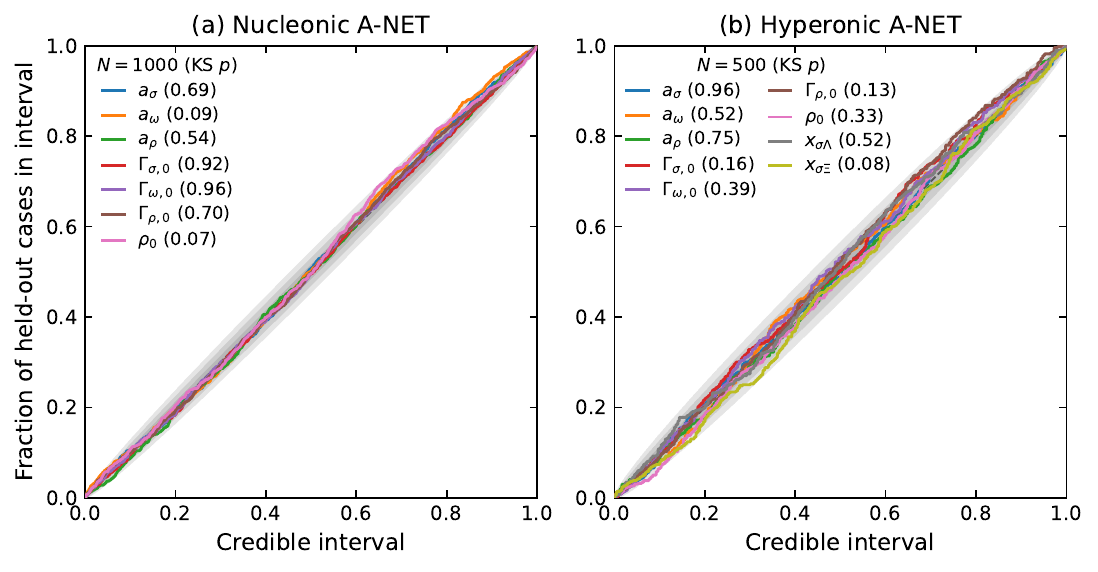}
  \caption{Marginal calibration (P--P) of the A-NET
  proposals on held-out simulated data for \textbf{(a)} the nucleonic DDB
  model ($1{,}000$ cases) and \textbf{(b)} the
  nine-parameter network of the hyperonic DDB$\Lambda\Xi^-$ A-NET ($500$
  cases). Grey bands: 68\%, 95\%, and 99.7\% ranges for a
  perfectly calibrated posterior; legend: Kolmogorov--Smirnov $p$
  values~\cite{Kolmogorov:1933,Smirnov:1948}.}
  \label{fig:anet-pp}
\end{resultfigurewide}

\section{Numerical convergence and robustness}
\label{app:numerics}

\begin{table}[tbp]
  \centering
  \caption{Largest changes over the eight configurations (four for the NICER
  prior shapes) when the likelihood
  construction is modified. $\Delta R_{1.4}$: over the 5th, 50th, and 95th
  percentiles; $\Delta R_{\rm band}$: over the 90\% band edges between $1.0$ and
  $2.0\,M_\odot$; $\Delta\Lambda_{1.4}$: median.}
  \label{tab:numerics}
  \footnotesize
  \setlength{\tabcolsep}{2.5pt}
  \begin{tabular}{@{}lcccc@{}}
    \toprule
    Modification & $\Delta R_{1.4}$ & $\Delta R_{\rm band}$ & $\Delta\Lambda_{1.4}$ & $\Delta\log\mathcal Z$ \\
     & (km) & (km) & & \\
    \midrule
    Finer quadrature and grids & $0.014$ & $0.028$ & $1.7$ & $0.15$ \\
    Other NICER KDE subsets & $0.023$ & $0.023$ & $3.3$ & $0.09$ \\
    Other GW KDE subsets & $0.017$ & $0.029$ & $5.0$ & $0.13$ \\
    NICER bandwidth ($\pm25\%$) & $0.004$ & $0.009$ & $0.7$ & $0.02$ \\
    GW bandwidth ($\times0.8$, $1.25$) & $0.031$ & $0.031$ & $3.5$ & $0.06$ \\
    Chirp mass integrated over & $0.063$ & $0.064$ & $15$ & $0.22$ \\
    $80$ stars per sequence & $0.003$ & $0.011$ & $0.7$ & $0.004$ \\
    GW prior divided out & $0.005$ & $0.005$ & $0.7$ & $0.41$ \\
    NICER prior shapes divided out & $0.018$ & $0.018$ & $3.1$ & $5.5$ \\
    \bottomrule
  \end{tabular}
\end{table}
\begin{table}[tbp]
  \centering
  \caption{Nucleonic A1 A-NET+IS with $64$, $256$, and $1024$ Heun steps and the
  same base draws: percentiles of $R_{1.4}$ (km) and the median
  $\Lambda_{1.4}$. Bootstrap Monte Carlo errors of the medians are given in
  parentheses; those of the 5th and 95th percentiles are $0.02$--$0.08$~km.
  The evidences use the density accumulated along the trajectory; the
  exact-map check is described in the text.}
  \label{tab:heun}
  \footnotesize
  \setlength{\tabcolsep}{2pt}
  \begin{tabular}{@{}lcccccc@{}}
    \toprule
    Run & $N_{\rm eff}$ & $\log\mathcal Z$ & $R_{1.4}^{5\%}$ & $R_{1.4}^{50\%}$ & $R_{1.4}^{95\%}$ & $\Lambda_{1.4}$ \\
    \midrule
    64 steps & $681$ & $-26.472$ & $11.71$ & $12.435\,(18)$ & $13.03$ & $440\,(4)$ \\
    256 steps & $1130$ & $-25.665$ & $11.71$ & $12.429\,(15)$ & $13.02$ & $438\,(3)$ \\
    1024 steps & $505$ & $-25.606$ & $11.70$ & $12.421\,(19)$ & $13.02$ & $436\,(4)$ \\
    \bottomrule
  \end{tabular}
\end{table}
\begin{table}[tbp]
  \centering
  \caption{Independent A-NET importance-sampling estimates (IS) of the evidences
  in Table~\ref{tab:en-ultranest}, with the EN--IS separation in units of the EN
  uncertainty.}
  \label{tab:en-is}
  \footnotesize
  \setlength{\tabcolsep}{2.5pt}
  \begin{tabular}{@{}lcccc@{}}
    \toprule
     & \multicolumn{2}{c}{Nucleonic DDB} & \multicolumn{2}{c}{Hyperonic DDB$\Lambda\Xi^-$} \\
    Configuration & IS & $\Delta/\sigma_{\rm EN}$ & IS & $\Delta/\sigma_{\rm EN}$ \\
    \midrule
    A1 base & $-25.606\pm0.042$ & $-0.15$ & $-28.727\pm0.012$ & $+0.01$ \\
    $K_0=200$ & $-25.540\pm0.057$ & $-0.79$ & $-29.136\pm0.015$ & $+0.16$ \\
    $K_0=260$ & $-25.825\pm0.044$ & $-0.53$ & $-28.583\pm0.014$ & $-0.34$ \\
    $J_{\mathrm{sym}}=29$ & $-25.559\pm0.028$ & $+0.13$ & $-28.695\pm0.012$ & $+0.04$ \\
    $J_{\mathrm{sym}}=36$ & $-25.660\pm0.140$ & $-0.53$ & $-29.029\pm0.015$ & $+0.96$ \\
    J0614$-$3329 & $-27.152\pm0.046$ & $-0.11$ & $-31.371\pm0.020$ & $-0.34$ \\
    J1231$-$1411 & $-26.363\pm0.027$ & $+0.33$ & $-29.225\pm0.011$ & $+0.89$ \\
    J1614$-$2230 & $-25.572\pm0.030$ & $-0.07$ & $-28.762\pm0.012$ & $+0.15$ \\
    \bottomrule
  \end{tabular}
\end{table}

\emph{Likelihood construction.} We construct the NICER densities using a
bandwidth factor of $0.08$ on a $150\times150$ grid spanning the 1st--99th
sample percentiles. Outside this grid, each density is set to $10^{-6}$ of its
maximum. Sources containing more than $2\times10^4$ samples are subsampled to
this size with probabilities proportional to their weights.
Equation~\eqref{eq:nicer-like} uses $40$ masses and Eq.~\eqref{eq:gw-like}
uses $20$ mass-ratio values, while the GW density uses Scott's bandwidth and
$4\times10^3$ randomly selected samples.

We test these choices for A1 and the three source substitutions,
PSR~J0614$-$3329, J1231$-$1411, and J1614$-$2230, in both compositions. Each
configuration contains $2979$--$29{,}336$ posterior samples with exact stellar
branches. We reweight these samples by the ratio of the modified to the
production likelihood, so sampling noise largely cancels when measuring the
shifts. Table~\ref{tab:numerics} reports tests using $400$ and $2000$ masses, a
$400\times400$ KDE grid, $200$ mass-ratio values, and different random KDE
subsets. We also vary the NICER bandwidth to $0.06$ and $0.10$, scale the GW
bandwidth by $0.8$ and $1.25$, double the number of stars along each stable
sequence, and divide out the analysis priors.

Grid and quadrature refinements change nucleonic $R_{1.4}$ by less than
$0.01$~km. The largest radius changes occur for hyperonic PSR~J1614$-$2230.
Changes in $\log\mathcal Z$ remain at most $0.05$ except for
PSR~J1231$-$1411, where they reach $0.15$ and $0.12$ for nucleonic and
hyperonic matter because its mass posterior lies near the $1\,M_\odot$
integration boundary. Dividing out the analysis priors mainly changes the
evidence normalization, producing an almost uniform shift in $\log\mathcal Z$.

The largest sensitivity comes from the fixed chirp mass in
Eq.~\eqref{eq:gw-like}. Integrating over it with a flat prior raises the median
$R_{1.4}$ by $0.01$--$0.05$~km and the median $\Lambda_{1.4}$ by $4$--$15$
because chirp mass and tidal deformability are correlated in the GW samples.
These tests characterize the systematic uncertainty of the common likelihood
used by all methods. The A1 log Bayes factor changes by at most $0.07$ across
these variants, except for chirp-mass integration, which lowers it by $0.24$.
Separately, removing the maximum-mass factor of Sec.~\ref{sec:method:like}
lowers the median $M_{\max}$ by $0.02$ and $0.04\,M_\odot$, and the median
$R_{1.4}$ by $0.01$ and $0.06$~km, for nucleonic and hyperonic matter,
respectively. The A1 log Bayes factor decreases by $0.52$, with nucleonic matter
still favored.

\emph{Flow integration.} We repeat the nucleonic A1 A-NET+IS analysis on CPU
with $64$, $256$, and $1024$ Heun steps using the same base draws. The
$1024$-step run uses only the first $4000$ draws. With the
trajectory-integrated density, the log evidence changes by $0.87$ between $64$
and $1024$ steps (Table~\ref{tab:heun}). We trace this shift to the difference
between that density and the exact density of the discrete Heun map,
calculated from the Jacobian of every step.

At $64$ steps, the difference in $\log q$ has a posterior-weighted mean of
$0.87$ and a standard deviation of $0.10$. Using the exact density changes the
ESS by $0.4\%$, the $R_{1.4}$ percentiles by at most $0.007$~km, and the median
$\Lambda_{1.4}$ by $1.7$. It gives log evidences of $-25.598$, $-25.599$, and
$-25.602$ for $64$, $256$, and $1024$ steps, respectively. The constant part of
the density offset cancels when the weights are normalized, while the
remaining posterior changes fall within the bootstrap uncertainties. For the
$1024$-step hyperonic heads, the offsets are $0.0013$--$0.0019$ with a standard
deviation of $0.0004$. Their effect on the mixture weights is negligible at the
reported Monte Carlo precision in the tested case.

\emph{Repeated runs and Monte Carlo errors.} Three independently seeded
nucleonic UltraNest runs at $J_{\rm sym}=36$~MeV agree within $0.014$~km in the
$R_{1.4}$ percentiles, $0.003\,M_\odot$ in median $M_{\max}$, and $4$ in median
$\Lambda_{1.4}$. For A1, bootstrap errors in the $R_{1.4}$ percentiles are
$0.007$--$0.013$~km for UltraNest and $0.022$--$0.029$~km for A-NET+IS, whose
ESS is about $8.5$ times smaller. The differences reported in
Sec.~\ref{sec:results} are comparable to these errors and consistent with Monte
Carlo noise. For hyperonic A1, all nine parameter medians from A-NET+IS agree
with UltraNest within $2\%$ of its 90\% credible-interval half-width. This is
comparable to the variation between two UltraNest seeds. The largest
corresponding difference for TSNPE+MIS is $6\%$.

\emph{EN uncertainty.} We first combine the scatter between networks, the Monte
Carlo error of the bank importance-sampling estimate, and the calibration error
as
\[
\sigma=\sqrt{\sigma_{\rm net}^{2}+\sigma_{\rm bank}^{2}+\sigma_{\rm cal}^{2}}.
\]
For nuclear updates, the calibration term is the root-mean-square error of the
second network output across the five nuclear settings. For held-out sources,
it combines the test errors on held-out synthetic and real baseline sources in
quadrature.

For each of the eight nucleonic configurations, we take the larger of this
combined uncertainty and the uncertainty from the EN analysis performed before
the bank extension. The earlier uncertainty is larger in all eight cases. For
held-out sources, we then add the difference $\Delta$ between the network and
bank estimates in quadrature. The released package contains all uncertainty
components and the script that combines them.

\emph{Independent evidence test.} The A-NET+IS draws provide evidence estimates
independent of the EN through $\hat{\mathcal Z}=N^{-1}\sum_i\widetilde w_i$.
Table~\ref{tab:en-is} compares the EN results with estimates from stored
$1024$-step hyperonic runs and new $1024$-step nucleonic runs containing $4000$
draws. The Monte Carlo errors in $\log\mathcal Z$ range from $0.011$ to $0.14$.
For hyperonic matter, these errors are $0.011$--$0.020$, and the EN differences
are at most $0.11$, with five of the eight below $0.03$.

\renewcommand{\bibfont}{\fontsize{8}{8.5}\selectfont}
\bibliography{refs}

\end{document}